\documentclass[a4paper,11pt]{article} 
\usepackage[dvips]{graphicx}
\usepackage{float}   
\usepackage{epsfig}
\usepackage{tabularx,longtable}
\title{Thermodynamic and kinetic aspects of RNA pulling experiments}
\author{M. Ma\~{n}osas and F. Ritort\\
Departament de Fisica Fonamental,
Universitat de Barcelona,\\
Diagonal 647, 08028 Barcelona, Spain}

\newcommand{\be}{\begin{equation}}
\newcommand{\ee}{\end{equation}}
\newcommand{\ba}{\begin{eqnarray}}
\newcommand{\ea}{\end{eqnarray}}
\newcommand{\eq}[1]{~(\ref{#1})}
\newcommand{\eqq}[2]{~(\ref{#1},\ref{#2})}
\newcommand{\eqqq}[3]{~(\ref{#1},\ref{#2},\ref{#3})}

\begin{document}   % End of preamble and beginning of text.

\bibliographystyle{plain}

\maketitle                 % Produces the title.
				
\nocite{*}

\abstract{ Recent single-molecule pulling experiments 
have shown how it is possible
to manipulate RNA molecules using optical tweezers force microscopy.
 We investigate a minimal model for the
 experimental setup which includes a RNA molecule
 connected to two polymers (handles) and a bead, trapped 
in the optical potential, attached to one of the handles. 
Initially, we focus on small single-domain RNA molecules 
which unfold in a cooperative way.
The model
qualitatively reproduces the experimental results and allow us to
investigate the influence of the bead and handles on the unfolding
reaction. A main ingredient of our model is to consider
 the appropriate statistical ensemble and the corresponding
thermodynamic potential describing thermal fluctuations in the
system. We then investigate several questions relevant to extract
thermodynamic information from the experimental data 
. Next, we study the kinetics using a dynamical model.
 Finally, we address the more general problem of a
multidomain RNA molecule with $Mg^{2+}$-tertiary contacts that unfolds
in a sequential way and propose 
techniques to analyze the breakage force data in order to obtain the
reliable kinetics parameters that characterize each domain.}

\section{Introduction}
\label{intro}
The RNA molecule plays a central role in molecular biology showing an
enzymatic function during the translation and splicing processes
\cite{b1,b2}. Experiments based on the manipulation of
single-biomolecules, such as laser tweezers force microscopy, allow
scientists to investigate their mechanical properties. These give
information about the structure, stability and the interactions involved
in the formation of such structures \cite{b3,b4,b5,b6,b7,b8,b9}.  In
these experiments mechanical force is applied to the ends of a RNA
molecule. The molecule is then pulled \cite{b10,b11} until a value of
the force is reached such that the molecule unfolds.  If the pulling
process is reversed then the molecule refolds again. In these
experiments the force exerted upon the system is recorded as a function
of the end-to-end distance giving the so-called force-extension curve
(FEC).  The nature of this unfolding-refolding process is stochastic and
therefore the values of the force at which the molecule unfolds-refolds
change from experiment to experiment.  Sometimes, as in the case of
presence of $Mg^{2+}$-tertiary contacts, it is not possible to pull the
molecule in quasi-static conditions because the relaxation time is too
large for the experimental possibilities which are largely limited due
to the presence of strong drift effects in the machine. Therefore, during
the pulling process, the molecule is driven to a non-equilibrium state
which is characterized by strong irreversibility effects.  The study of
this pulling process might be useful to understand many biological
processes where biomolecules are unfolded under locally applied force,
for example when the mRNA goes through the ribosome during the
translation process.

To manipulate a RNA molecule some synthesized polymers typically several
hundred nanometers long (called handles) have to be chemically linked to the
extremes of the RNA molecule. A polysterene bead is then chemically
attached to the end of one of these handles and used to measure the
force by reading its position inside the optical trap. These
additional elements (bead and handles) are an inseparable part of any
pulling experiment and they have an influence on the unfolding
process. To characterize the thermal behavior of the pulled global
system (bead, handles plus RNA molecule) it is important to identify
the proper control parameter. This is an essential step towards the
modelization of the experiment and has several consequences. 
For instance, the force acting on the extremes of the RNA molecule cannot
be externally controlled but fluctuates and its mean value depends in a 
non-linear way on the value of the control parameter.  The control parameter
determines the relevant thermodynamic potential defining the
equilibrium state of the global system as well as the magnitude of the
fluctuations around that state.  A proper inclusion of these parts is
necessary to accurately interpret the experimental data. Another
important point of the work is the model for the RNA molecule. We 
consider the RNA molecule to be composed by different domains, each one showing
cooperative unfolding. Each domain is then modeled as a two-states
system: the unfolded state (UF) and the folded one (F), which are 
separated by a kinetic barrier. A main effort throughout this paper is
to present in the most clear way the appropriate theoretical frame to
understand pulling experiments leaving aside further additional
complications, nevertheless important, such as a detailed response of
the optical tweezers machine or the microscopic structure of the RNA
molecule.

The goal of this paper is twofold: (i) we
 show how to build a minimal model aiming to reproduce the
experimental setup including all the aforementioned elements
(bead, handles and the RNA molecule) and quantitatively
reproducing various experimental results; (ii) we show 
how to analyze experimental data extracted from both
quasi-static and out-of-equilibrium pulling experiments in order to
obtain thermodynamic and kinetic information about the unfolding
reaction.

The paper is divided into three main parts. In the first part of the paper
(Sections~\ref{setup},\ref{twostates},\ref{parts}) we describe the
model for the experimental setup (Sec.~\ref{setup}) and introduce the
ensemble that is relevant to model the pulling experiment
(Sec.~\ref{ens}). In Sec.~\ref{twostates} we describe the two-states
model convenient to reproduce the cooperative unfolding of the RNA
molecule and in Sec.~\ref{parts} we describe the models used for the
bead and handles.  In the second part of the paper
(Sections~\ref{thermo},\ref{simulation}) we analyze the
unfolding-refolding behavior of a cooperative two-states RNA molecule
in a pulling experiment for both equilibrium and non-equilibrium
regimes. For the equilibrium regime, we compute the partition function
in the ensemble that is experimentally relevant, and derive an
expression for the quasi-static work exerted upon the system as the
molecules unfolds. This expression relates the work measured in a
quasi-static pulling process to the difference of free energy between
the F and UF states at zero force $\Delta G^0$. We analyze in
detail the different thermodynamic contributions to the total work,
the influence of the parameters describing bead and handles on the FEC,
and obtain expressions for the force at the midpoint of the
transition. For the non-equilibrium behavior we investigate in detail
the fraction of molecules that unfold (refold) more than once during
the unfolding (refolding) path, which is a quantity amenable to
experimental checks. We find that this fraction is related to the mean
dissipated work exerted upon the system, which gives us a way to
extract the reversible work in non-equilibrium processes just by
measuring the total work.  We also identify an interesting symmetry
property relating these fractions for the forward and reverse
processes.  To endorse most of our theoretical results we also
consider a simulation of a pulling experiment that allow us to obtain the
characteristic FEC, either in a situation
where the transition occurs in equilibrium or in a situation where it
does not.  In the third part of the paper (Sec.~\ref{domains}), we
address the unfolding behavior of complex RNA molecules with more than
one folded-domain and in the presence of $Mg^{2+}$-dependent
barriers. In this case, the refolding is not observed at the
experimental conditions, and the distribution of the breakage force is
a first order Markov process \cite{b12,b13}. We focus our attention in
the specific case of RNA molecules where domains unfold in a
sequential fashion according to a reproducible path. This unfolding
mechanism is generally a consequence of the topological connectivity
of the different parts of the molecule and of the blockade of the force
induced by the most external tertiary contacts on the interior
domains. We model the molecule as a series of domains, each
represented by a two-states system, and we compute the distribution of
breakage force for each domain. We propose several methods of
analyzing the breakage force data in order to achieve reliable values
for the height and position of the barrier of each domain.  In Sec.
\ref{conclusions} we present the conclusions. Five appendixes are devoted to describe some analytical calculations.
    
\section{Model for the experimental setup}
\label{setup}
\begin{figure}[H]
\begin{center}
\includegraphics[scale=0.35]{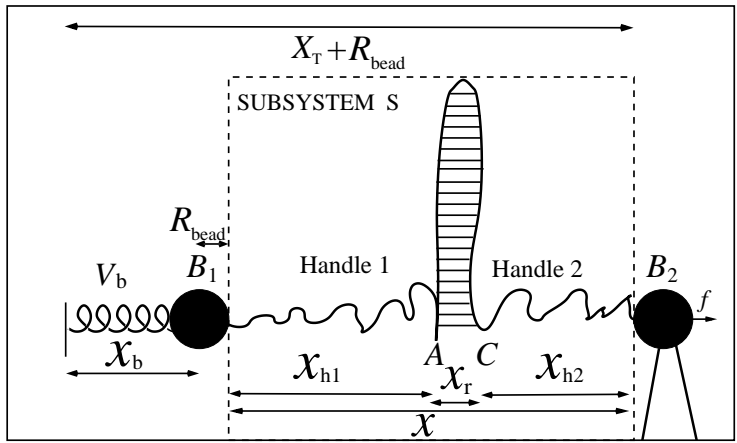}
\caption{\small{Schematic picture of the model for the experimental setup 
in a RNA pulling experiment as described in the text.
 We show the configurational variables of the
system $x_b,x_r,x_{h_1},x_{h_2}$ which are the projections of the
end-to-end distance of each element along the reaction coordinate axis
(i.e. the axis along which the force is applied). The potential $V_b(x_{b})$ is well described by an harmonic potential of one-dimensional spring with rest position at $x_{b}=0$.}}
\label{f1}
\end{center}
\end{figure}

We consider a minimal model in order to reproduce the experimental
setup of a pulling experiment carried out using laser tweezers force microscopy
\cite{b10,b13bi}. The model (Fig.~\ref{f1}) is composed by a small RNA molecule connected
to two polymers called handles~\footnote{\label{foot1}These are hybrids 
of DNA and RNA rather than
single stranded DNA or RNA polymers in order to avoid the formation of 
secondary structures} which are used to attach the small RNA molecule to two
beads at each end. One bead is confined in the optical trap generated by
the laser beams, the other is held fixed to the tip of a micropipette
 by air suction.
The whole system consists of a chain of connected elements. Starting
from the left side of the chain there is a bead ($B_1$) of radius
$R_{\rm bead}$ that is trapped in the laser tweezers potential,
$V_{b}(x)$~\footnote{\label{foot2}Although the trap potential should be defined in
the three-dimensional space $V_{b}(\vec{x})$ we will consider
$V_{b}(x)$ as the potential of mean-force projected along the reaction
coordinate axis. This approximation is very accurate as fluctuations
along the $y,z$ directions are assumed to hardly affect the unfolding
behavior of the RNA molecule.}. We use the position of this bead
$B_1$ to read the force acting on the system in the same way as the
needle of a 'manometer' is used to read the pressure exerted by a gas on
the walls of a container~\footnote{\label{foot3}This is not the way the force is
usually measured in dual beam optical tweezers where two
photodetectors located at opposite sides of the chamber are used to
collect the total amount of deflected light which is then converted
into force after calibration of the machine.}. The second element is a
handle (handle 1) with one end specifically attached to the bead $B_1$
and the other end attached to the RNA molecule at the point
$A$. The second handle (handle 2) has one end specifically attached to
the RNA molecule at the point $C$.  The other end is specifically
attached to the bead $B_2$, fixed to the tip of a
micropipette. The molecule is pulled by moving the micropipette along
the $x$ direction . The configurational variables of this simplified
system are taken as the projections of the end-to-end distances of
each element along the force axis: $x_{h_{1}}=\overline{B_1 A}-R_{\rm
bead},~x_{h_2}=\overline{C B_2}-R_{\rm bead}$ for the distances of the
handles, $x_{r}=\overline{AC}$ for the RNA end-to-end distance and
$x_{b}$ for the position of the bead $B_1$ in the trap. The force $f$
is measured by reading the position $x_{b}$ of the bead $B_1$:
\be f=\left|\frac{dV_{b}(y)}{dy}\right|_{y=x_{b}}.
\label{fs}
\ee 
We define the subsystem $S$ as that composed by the two handles and
the small RNA molecule. The end-to-end distance for the subsystem $S$
is then given by $x=x_{h_1}+x_{h_2}+x_r$ (Fig.~\ref{f1}). The
total distance between the center of the trap and the tip of the
micropipette is given by $X_{T}+R_{\rm bead}=x_{b}+x+R_{\rm
bead}$. Pulling experiments give FECs, $f(x)$, corresponding to the
force exerted on the chain \eq{fs} measured through the position of the bead
$B_1$ as a function of the end-to-end distance of subsystem
$S$.

\subsection{Ensembles}
\label{ens}
Experimentally it is possible to consider two different
ensembles depending on which variable is used as the externally
imposed non-fluctuating parameter~\footnote{\label{foot4}The existence of an
external non-fluctuating parameter is required to have a well defined
equilibrium state.}.
\begin{itemize}
\item{\bf Mixed ensemble:} The total distance between the center of
the trap and the tip of the micropipette is held fixed, hence $X_{T}$
is the externally controlled parameter. In this ensemble there are
fluctuations in $x$ and $f$ given by \cite{b14,b15},
\begin{eqnarray}
\langle \delta x^{2}\rangle=\frac{k_{B}T}{k_{x}(X_{T})+k_{b}(X_{T})}~,\;\;\; \langle \delta f^{2}\rangle =\frac{k_{B}Tk_{b}^{2}}{k_{x}(X_{T})+k_{b}(X_{T})}~,\nonumber\\ 
{\rm {with}} ~~k_{x}(X_{T})=\frac{d\langle f\rangle}{d\langle x\rangle}\Big{|}_{X_{T}}, ~~ k_{b}(X_{T})=\frac{d\langle f\rangle}{d\langle x_{b}\rangle}\Big{|}_{X_{T}},
\label{flumix}
\end{eqnarray}
where $\langle...\rangle$ stands for thermal average, $k_{B}$ is the
Boltzmann constant, $T$ is the temperature of the bath, $k_{b}(X_{T})$
is the stiffness of the optical trap and $k_{x}(X_{T})$ is the
effective rigidity corresponding to the subsystem $S$. The latter is
determined by the serial compliance
\be
k_{x}(X_{T})=\Big{[}\frac{1}{k_{h_{1}}(X_{T})}+\frac{1}{k_{h_{2}}(X_{T})}+\frac{1}{k_{r}(X_{T})}\Big{]}^{-1},
\label{kx}
\ee
where $k_{h_{i}}$ ($i=1,2$) and $k_{r}$ are the rigidities of the
handles 1, 2 and the RNA respectively. These rigidities are $X_{T}$
dependent and so are the fluctuations \eq{flumix}.

\item{\bf Force ensemble:} In this case a piezo actuator controls the force (and
therefore the position of the bead $B_1$). In this ensemble $X_{T}$ and
$x$ are fluctuating variables, $\langle \delta X_{T}^{2}\rangle
=\langle \delta x^{2}\rangle=k_{B}T/k_{x}(f)$, where
$k_{x}(f)$ is the stiffness of the subsystem $S$ when the force is held fixed, $k_{x}(f)=\Big{[}\frac{d\langle x\rangle}{df}\Big{]}^{-1}$. 
\end{itemize}
Most of the theoretical work for the denaturation of RNA in 
pulling experiments
 considers the force ensemble. 
However, it is experimentally very difficult to work in the force ensemble
where either the force or the variable $x_{b}$ must be
controlled. Indeed, for $X_{T}$ to fluctuate the center of the trap
must also fluctuate to compensate for the fluctuations in the
force. It is difficult to imagine how to experimentally implement such
ensemble.  Therefore the most natural ensemble is that where $X_T$ is
constant. Indeed this is the ensemble most relevant for the
experiments and therefore we will work in the mixed ensemble throughout this
article.
   
\section{Two-states model for a single RNA domain under the effect of an external mechanical force}
\label{twostates}
The unfolding of some biomolecules under the effect of a mechanical
 force is a highly cooperative process that can be qualitatively
 described by a two-states model. The two-states model has a long
 tradition in physics and has been applied previously by several
 authors in order to explain the unfolding behavior of single domains
 of proteins and RNA hairpins
 \cite{b10,b16,b17,b18,b19,b20}. Recently, it has been shown how such a
 simple phenomenological description, with Kramer transitions-rates,
 does not fully reproduce the kinetics observed in pulling
 experiments of the protein Titin, and more realistic descriptions
have been proposed~\cite{HumSza03}.

\begin{figure}[H]
\begin{minipage}{7cm}
\begin{center}
%FEC1r.eps
\includegraphics[height=4cm]{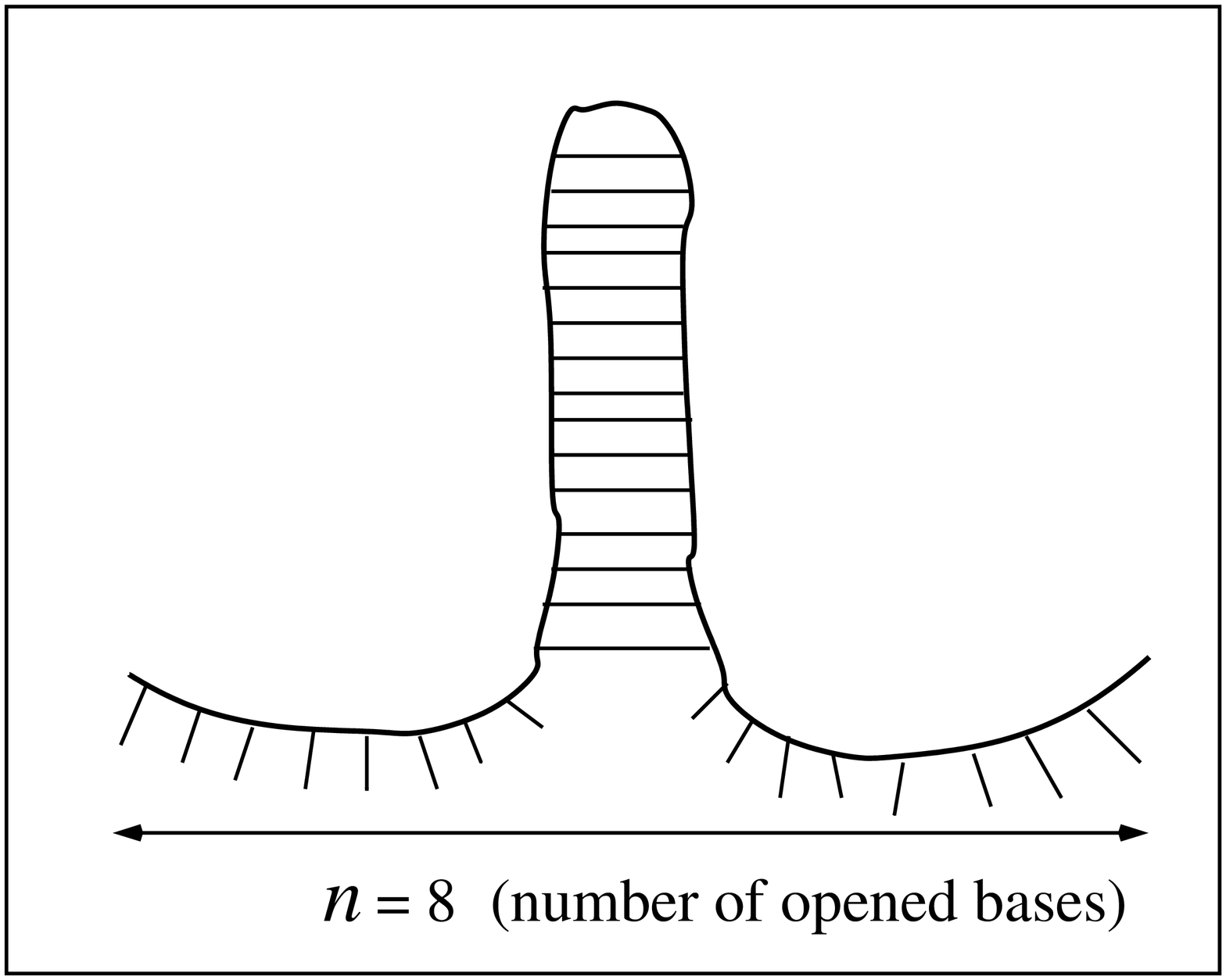}\\
(a)
\end{center}
\end{minipage}
\begin{minipage}{7cm}
\begin{center}
%FEC1p.eps
\includegraphics[height=5cm]{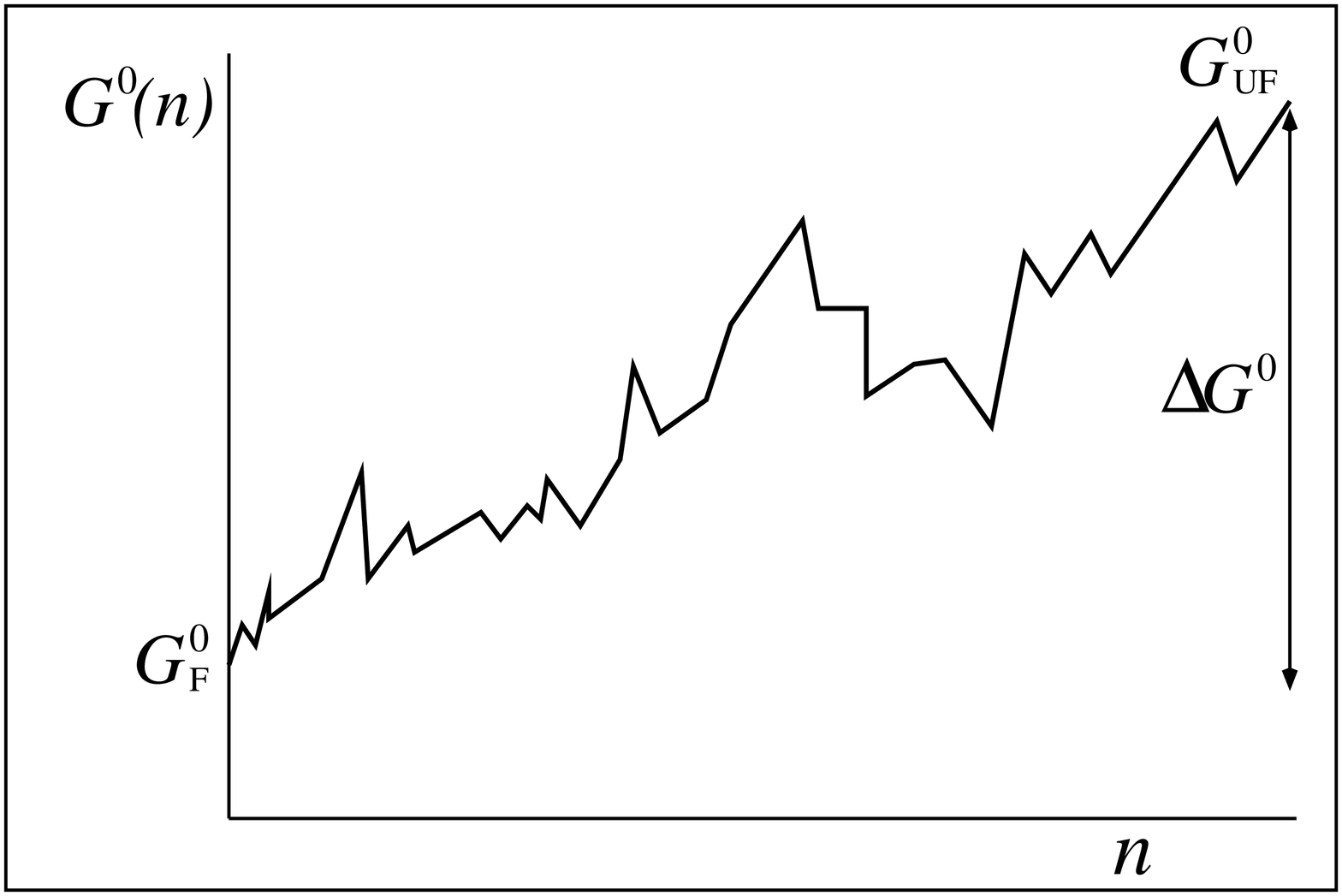}\\
(b)
\end{center}
\end{minipage}
\caption{\small{Schematic representation of (a) a RNA hairpin with
$n=8$ bases opened. (b) the free
energy landscape for a single RNA hairpin at zero force as a function of
the number of denaturated bases $n$ at $T<T_{\rm melting}$ (melting temperature of the RNA)  and normal
ionic conditions. In this situation the stable state is the folded one
with $n=0$.}}
\label{f2}
\end{figure}

 Let us consider an RNA molecule isolated from the rest of the system
 in equilibrium at constant temperature, pressure and zero force. In
 the simplest description both states (hereafter denoted by
 UF -unfolded- and F -folded-) are characterized by their Gibbs free
 energy $G^{0}_{\rm UF}$ and $G^{0}_{\rm F}$ respectively and the RNA
 molecule occupies each state with a probability given by the
 Boltzmann distribution. In a more refined description the molecule
 can also occupy intermediate configurations depending on the number
 $n$ of the first-opened, or denaturated, bases (
 Fig.~\ref{f2} (a))~\cite{CocMonMar}. The F and UF states correspond then to
 the RNA configuration with $n=0$ and $n=N$ bases opened, where $N$ is the
 total number of pair of bases of the molecule. The free energy
 landscape is described by a function $G^0(n)$ which characterizes the
 probability of a hairpin having the first $n$ bases opened (Fig.~\ref{f2} (b)). This
 description excludes the existence of other {\em breathing}
 intermediate configurations that might be relevant for the unfolding
 reaction~\cite{Marinarietal}.

\begin{figure}[H]
\begin{minipage}{7cm}
\begin{center}
%FEC1r.eps
\includegraphics[height=5cm]{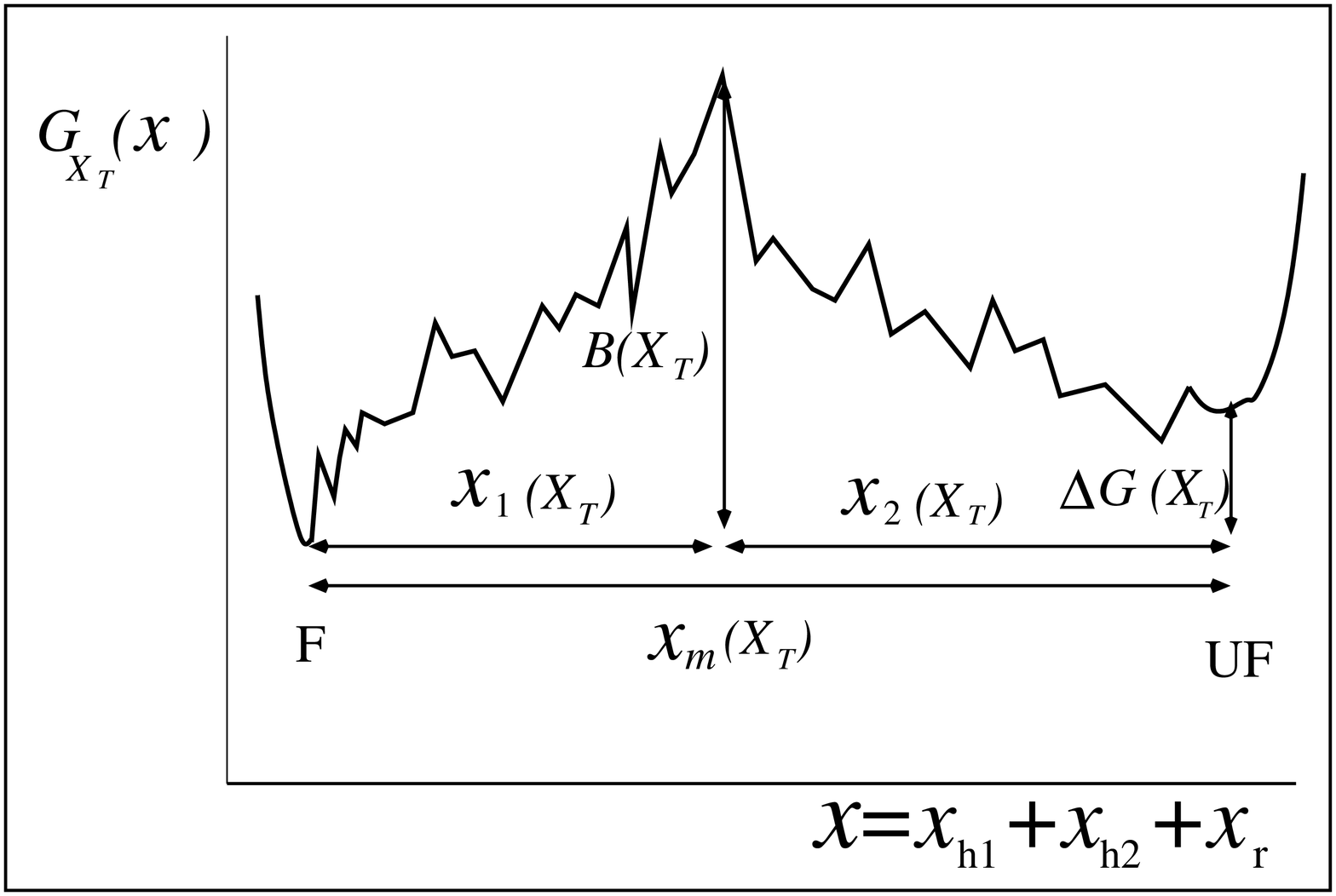}\\
(a)
\end{center}
\end{minipage}
\begin{minipage}{7cm}
\begin{center}
%FEC1p.eps
\includegraphics[height=4cm]{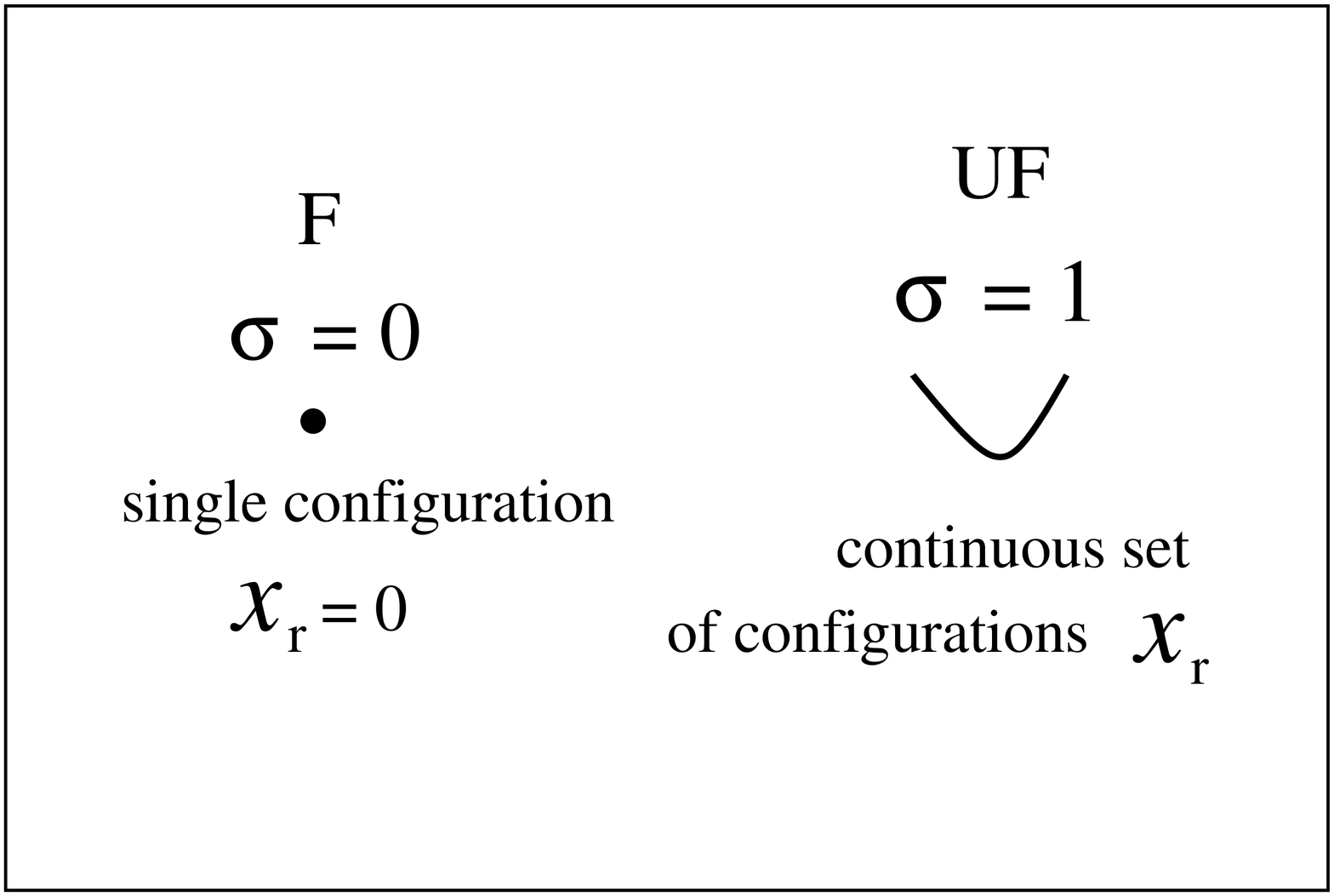}\\
(b)
\end{center}
\end{minipage}
\caption{\small{ (a) Schematic representation of the  free energy
landscape $G_{X_{T}}(x)$ for the whole system at $T<T_{\rm melting}$,
for $X_{T}<X_{T}^{c}$ (where $ X_{T}^{c}$ is the value of $X_{T}$ where both
states F and UF are equiprobable) and normal ionic conditions. In this
situation the stable state is still the folded one. We also show all the
parameters characterizing the two states model. (b) Schematic
representation of the relevant configurations in the F and UF states
along the reaction coordinate $x_{r}$. We consider the F state
to be characterized by a single configuration $x_{r}=0$ and the UF
state by a continuous set of values of $x_{r}$. We use the label $\sigma=0$ for the F state and $\sigma=1$ for the UF state.}}
\label{f3}
\end{figure}

When an externally controlled force $f$ is applied to the ends of the
RNA molecule the adequate thermodynamic potential to consider is the
Legendre transform of the Gibbs free energy $G'(n)=G^0(n)-fx_{r}(n)$ \cite{b21}. 
The free
energy landscape $G'$ is then tilted along the reaction coordinate
$x_r$, which is the projection of the end-to-end distance of the
molecule in the axis force and explicitly depends on the number of
opened bases $n$. Since we work in the ensemble where neither $f$ nor
$x_r$ are control parameters the non-fluctuating
parameter $X_T$ determines the adequate thermodynamic potential 
$G_{X_T}$. The free
energy $G_{X_T}$ of the system shown in Fig.~\ref{f1} is a
potential of a mean force that characterizes the equilibrium
state of the whole system, including the handles, the bead and the RNA molecule, at fixed value of $X_T$. In order to build
the model is useful to represent the free energy $G_{X_T}$, as a 
function of the end-to-end distance of the subsystem $S$, as shown in
Fig.~\ref{f3} (a). This picture tells us about the probability  $p_{X_T}(x)$ of
finding the subsystem $S$ at a given value of its end-to-end distance
$x$ for a fixed value of $X_T$, $p_{X_T}(x)\propto \exp(-\beta
G_{X_T}(x))$, where $\beta=1/k_{B}T$.

The free energy landscape $G_{X_T}(x)$ shows two pronounced minima
corresponding to the F and UF states (Fig.~\ref{f3}). The discrete variable $\sigma$
stands for the state of the domain: the value $\sigma=0$ denotes the F
state and $\sigma=1$ the UF state. The relative thermodynamic
stability of these states depends on the difference of free energy
between them, $\Delta G(X_{T})$. Moreover we will consider the
existence of an intermediate or transition state along the reaction
path from the F to the UF states and vice versa.  This transition
state is the intermediate RNA state with highest free energy 
connecting the F and the UF sate along the reaction path. It may correspond to
a RNA configuration where the first $n=n^{*}$ bases are
opened\footnote{\label{foot5}We stress that the shape of the free
energy landscape depends on $X_{T}$ as well as the location of the
barrier corresponding to the intermediate state. Therefore the value
of $n^{*}$ that characterizes the transition state is also $X_{T}$
dependent. There are experimental limitations to follow the folding
and unfolding of the molecule (hopping) given by the operational range
of frequencies of the instrument used. For instance, in \cite{b10} this
operational range was $0.05-20\rm Hz$, meaning that hopping events out
of this range were not observable.  Moreover, in pulling processes the
folding-unfolding reaction only occurs in a narrow range of values of 
the control parameter $X_{T}$ around $X_{T}^{c}$, otherwise 
the folding-unfolding relaxation time is too large.  Hence the study of
the folding-unfolding kinetics is restricted to the regime
$X_{T}\approx X_{T}^{c}$ and to the operational range of
frequencies. Therefore, as the transition state $n^{*}$ is only
relevant for the study of the kinetics, we assume $n^{*}$ independent of 
$X_{T}$, $n^{*}=n^{*}(X_{T}^{c})$.}. In the simplest scenario the
intermediate state can be assumed to have a very short lifetime, its
main effect is to hinder transitions from the F to the UF state and
back. In this scenario the transition state can be represented by an
activation barrier and this is the model we will adopt throughout the
paper.  The F and UF states are separated by a barrier of height
$B(X_{T})$ measured relative to the F state.  The barrier is located
at a distance $x_{1}(X_{T})$ from the F state and $x_{2}(X_{T})$ from the UF
state. The distance between the two states is
$x_{m}(X_{T})=x_{1}(X_{T})+x_{2}(X_{T})$. Because the rigidity of the
RNA molecule in the F state is very large we can assume this state to
be characterized by a single configuration corresponding to the value
$x_{r}=0$ of the reaction coordinate; fluctuations around this
configuration cost so much energy that they are highly improbable. The
RNA in the UF state has a finite rigidity, hence it is represented by
a set of configurations within a continuous range of values of $x_{r}$ 
(Fig.~\ref{f3} (b)).

\section{Modeling the different parts of the setup}
\label{parts}
In this section we specify how we model the different elements of 
the system: the
bead trapped in the optical tweezers potential, the two handles, and
the RNA.

\subsection{Model for optical tweezers: a bead matched to a spring}
\label{bead}
Typical experimental values for the trap stiffness and the diameter of
the beads are $k_{b}\approx 0.15-0.05\rm{pN/nm}$ and
$R_{\rm{bead}}\approx 1-3\rm{\mu m}$ respectively. We consider that
the bead follows a Langevin dynamics of an overdamped particle
(i.e. without inertial term)\footnote{\label{foot6}In (\ref{b1}) we are neglecting
the drag force felt by the bead (equal to $-\gamma v$) as the chamber
is moved (and the water dragged relative to the lab frame) at a
certain pulling speed $v=\frac{dX_{T}}{dt}$. For
the range of pulling speeds used in the experiments this contribution
is negligible, of the order of $0.1\rm pN$.} :
\be
{\gamma}\frac{dx_{b}}{dt}=F_{R}(x_{b})+{\xi}(t)~,
\label{b1}
\ee
where ${\gamma}$ is the friction coefficient and $F_{R}$ is the
resultant force applied to the bead. The stochastic term ${\xi}(t)$ is
a white noise with mean value $\langle {\xi}(t)\rangle=0$ and variance
$\langle {\xi}(t){\xi}(t')\rangle =2k_{B}T{\gamma}{\delta}(t-t')$.
The force $F_{R}$ has two contributions: one coming from the optical
trap potential, $f$, given by (\ref{fs}), and the other from the
subsystem $S$, $f_{x}$~\footnote{\label{foot7}This is also the force
exerted upon the subsystem $S$ for a given value of
$x=X_{T}-x_{b}$.}. Therefore $F_{R}=-f+f_{x}(x_{b})$.  The
experimental results \cite{b10,b11} show a linear dependence of the
force $f$ on $x$ along the transition (rip) where the variable $x$
refers to the subsystem $S$ (Fig.~\ref{f1}), hence we conclude
that the optical trap $V_{b}$ is well modeled by an harmonic potential
of stiffness $k_{b}$. We can then express the force $f$ measured
through the optical tweezers as $f=k_{b}x_{b}=k_{b}(X_{T}- x)$, where
we have used (\ref{fs}). In equilibrium the average force acting upon
the bead is zero $\langle F_{R}\rangle$, hence $\langle
f_{x}\rangle=k_{b}\langle x_{b}\rangle$. However $x_{b}$ fluctuates
and so both instantaneous forces $f$ and $f_{x}$ are not identical.
Doing an expansion around the equilibrium position of the bead,
$x_{eq}$~\footnote{\label{foot7b}We expand $f_{x}$ and $f$ around $x_{eq}$ 
keeping only the first term in $(x_{b}-x_{eq})$, i.e 
$f_{x}\approx \langle f_{x}\rangle+k_{x}(x_{eq}-x_{b})$, with $k_{x}$ 
given by \eq{kx}, and $f\approx \langle f\rangle+k_{b}(x_{b}-x_{eq})$. 
This approximation is valid in the regime where the force fluctuations are 
not big. Using that at equilibrium $\langle f\rangle=\langle f_{x}\rangle$ 
we obtain $F_{R}=-f+f_{x}\approx -(k_{x}+k_{b})(x_{b}-x_{eq})$.}, we obtain:
\be
{\gamma}\frac{dx_{b}}{dt}=-k_{R}(x_{b}-x_{eq})+{\xi}(t)~,
\label{b2}
\ee
where $k_{R}$ is the effective spring constant applied to the bead,
 $k_{R}=k_{x}+k_{b}$, with $k_{x}$ given by \eq{kx}. The relaxation
 time of the system (i.e the typical time during which the position of
 the bead decorrelates), $\tau_{b}$, is given by
 $\tau_{b}={\gamma}/k_{R}$. Applying the Stokes's law for the
 friction coefficient in water we obtain:
 $\gamma=6{\pi}R{\eta}\approx10^{-5}\rm pNs/nm$. The stiffness of the
 handles and the RNA are force dependent. Near the F-UF transition,
 typically these stiffness values are, at least, one order of
 magnitude bigger than $k_{b}$. Taking $k_{b}=0.1\rm pN/nm$, and
 $k_{h_{1}},k_{h_{2}},k_{\sigma}>\rm 1pN/nm$ we get
 $\tau_{b}<10^{-5}\rm s$~\footnote{\label{foot8}In absence of handles $k_{R}=k_{b}$
 and $\tau_{b}=10^{-3}\rm s$. Therefore $10^{-3}\rm s$ is the slowest
 relaxation time for the bead corresponding to the regime where the
 handles have practically no rigidity, $k_{x}\approx 0$, a situation
 only encountered at small forces (below 1pN approximately).}. By collecting data at
 frequencies smaller than $10^{5}\rm Hz$ we can guarantee that we will
 not have effects due to the bead's overdamping, hence the bead 
 relaxes quickly to its new equilibrium position at each step. This ensures
 that assuming an instantaneous relaxation of the bead position is
 enough to capture its overdamped dynamics.

\subsection{Polymer model for handles and single-stranded RNA}
\label{polymer}
The handles and the single-stranded RNA (ssRNA), the UF state of the RNA,
 are polymers that
typically measure $d\approx1-3 {\rm nm}$ in diameter and $L\approx20-500
{\rm nm}$ in length. As the bead has a much bigger size than the
polymers the friction coefficient (and therefore the relaxation time)
for the polymers is much smaller. This allows us (as we do for the bead) to only consider an instantaneous relaxation for the handles and the ssRNA. To describe the equilibrium behavior of the
handles and the ssRNA under the effect of an external force we use the
worm-like-chain (WLC) model \cite{b22}. This is described by a
Hamiltonian that includes a local bending term as well as the potential
energy of the polymer in the presence of the pulling
force. Parameterizing the polymer with the arc length $s$, the energy
function can be written as:
\be E_{WLC}=
\int_{0}^{L_{o}}\Big{[}\frac{k_{B}TP}{2}\Big{(}\frac{d\vec{t}(s)}{ds}\Big{)}^{2}-f\cos \theta(s)\Big{]}\,ds~,
\label{p1}
\ee
where $L_{o}$ is the contour length of the polymer, $\vec{t}(s)$ is the
 unit tangent vector along $s$, $\theta(s)$ is the angle between
 $\vec{t}(s)$ and the force axis, and $P$ is the persistence length
 defined as the typical distance over which $\vec{t}$-correlations decay to zero:
 $\langle \vec{t}(s)\vec{t}(s') \rangle\approx{e^{-\frac{\vert
 s-s'\vert}{P}}}$. The persistence length of a polymer depends on the
 ionic conditions \cite{b23}, and typical values are $50 {\rm nm}$ for
 double-stranded DNA (dsDNA) and $1 {\rm nm}$ for ssRNA. The thermodynamic properties of this
 model cannot be exactly computed, yet there are useful extrapolation
 formulas. Bustamante et al. \cite{b24} have proposed a simple expression
 for the force as a function of mean end-to-end distance of the polymer $x$,
\begin{equation}
f=\frac{k_{B}T}{P}\Big{[}\frac{1}{4(1-x/L_{o})}-1/4+x/L_{o}\Big{]}~.
\label{p2}
\end{equation}
Eq.(\ref{p2}) gives the exact solution as $x$ approaches either zero
or $L_{o}$ and is accurate at least up to $90\%$ in between.
Bouchiat et al. \cite{b25} have given an expression with an accuracy of
$99\%$ by adding to (\ref{p2}) a polynomial of seventh order. The WLC
model works well only at low forces, in the so called entropic regime
where the molecule behaves as an entropic spring. At high forces there
is an enthalpic correction due to the fact that the bonds are stretched
and the contour length $L_{o}$ increases. To incorporate this effect it
is then enough to replace $x/L_{o}$ by $x/L_{o}-f/E_{y}$ in \eq{p2},
 where $E_{y}$ is the Young modulus of the polymer
(typical values are $E_{y}\approx{500-1500}\rm{pN}$ for DNA-RNA molecules).\\
Throughout Secs.~\ref{thermo} and \ref{simulation}  we  analyze the
unfolding dynamics and thermodynamics of a single hairpin of RNA aiming to reproduce the
results obtained from a pulling experiment for the hairpin P5ab in 10mM $Mg^{2+}$ \cite{b10}. In Sec.~\ref{thermo} we use the partition function analysis to
individuate the different thermodynamic contributions to the total free
energy or reversible
work done upon the system. Next in Sec.~\ref{simulation} we do numerical simulations of a pulling experiment.

\section{Thermodynamic analysis}
\label{thermo}
In this section we use the tools of statistical mechanics to analyze
the thermodynamics of the system represented in Fig.~\ref{f1}. Most of
the analytical development is done in appendix~\ref{append_part} and
in Sec.~\ref{def} we give the main results. In Sec.~\ref{WT} we show
how to get the force-extension curve (FEC), the value of the force at
the midpoint of the F-UF transition $F^c$, and the different
contributions to the total reversible work coming from the
different elements of the system. In Sec.~\ref{Wrip} we derive an
expression that relates the reversible work exerted upon the subsystem
$S$ across the transition with the difference of free energy between
the F and UF states at zero force, $\Delta G^0$.  As this is an
experimentally measurable quantity this procedure provides a way to
estimate the unfolding free energy of the molecule, a quantity
biologically relevant as it determines the fate of biochemical
reactions.  Finally in Sec.~\ref{compWrip} we show how to apply these
relations to a specific example.
\subsection{Definitions}
\label{def}
In equilibrium the observables $x_{\alpha}$ and the conjugated forces
$f_{\alpha}$ with $\alpha=h_{1},h_{2},r,b$
(referring to the different elements, handle 1 and 2, RNA and bead
respectively) fluctuate. However, the thermodynamic free energy is only
a function of the mean values of these observables that we denote
by $\langle x_{\alpha}\rangle,~\langle f_{\alpha}\rangle$. A
representation of $\langle f_{\alpha}\rangle$ versus $\langle
x_{\alpha}\rangle$ gives what we call the thermodynamic force
extension curve (TFEC) for the element $\alpha$ in the mixed
ensemble. If $\alpha$ refers to the whole subsystem $S$ then the TFEC
corresponds to the usual force-extension recorded in RNA pulling
experiments, assuming that the pulling process is carried out
reversibly.  We can also define the restricted average $\langle{\cal
O}\rangle_{\sigma}(X_{T})$ as the mean value of the observable ${\cal
O}$ when the RNA molecule is in the state $\sigma$ for a fixed total
end-to-end distance $X_{T}$. From now on, all the dependencies of the
observables on the variable $X_T$ will not be explicitly written,
hence $\langle {\cal O}\rangle_{\sigma}(X_{T})\equiv \langle {\cal
O}\rangle_{\sigma}$.  In appendix~\ref{append_part} we derive an
expression for the partition function $Z(X_{T})$, corresponding to the
system schematically represented in Fig.~\ref{f1}. Applying the saddle
point technique, and separating the contributions that come from the F
($\sigma=0$) and UF ($\sigma=1$) states we get:
\begin{eqnarray}
Z(X_{T})&=&Z_0(X_{T})+Z_1(X_{T})~, \label{r1a} 
\end{eqnarray}
where
\begin{eqnarray}
Z_{0}(X_{T})\approx\exp\Bigl[-{\beta}\Big{(}W_{h_{1}}(\langle x_{h_{1}}\rangle_{0})+W_{h_{2}}(\langle x_{h_{2}}\rangle_{0})+V_{b}(\langle x_{b}\rangle_{0})\Big{)}\Bigr],\label{r1b}
\end{eqnarray}
\begin{eqnarray}
Z_{1}(X_{T})\approx\exp\Bigl[-{\beta}\Big{(}W_{h_{1}}(\langle x_{h_{1}}\rangle_{1})+W_{h_{2}}(\langle x_{h_{2}}\rangle_{1})+V_{b}(\langle x_{b}\rangle_{1})+{\Delta}G^{0}+W_{r}(\langle x_{r}\rangle _{1})\Big{)}\Bigr]~.
\label{r1c}
\end{eqnarray}
Here $V_{b}$ represents the optical trap potential and ${\Delta}G^{0}$
is the free energy difference between the F and the UF states at zero
force.  The function $W_{\alpha}(x)$ corresponds to the reversible
work performed by adiabatically stretching the element $\alpha$ from
$x_{\alpha}=0$ to $x_{\alpha}=x$ and it is given by
\begin{eqnarray}
W_{\alpha}(x_{\alpha})=\int_{0}^{x_{\alpha}}dx f_{\alpha} (x), \;{\rm with} \; \;\alpha=h_{1},h_{2},r~,  
\label{rr1}
\end{eqnarray}
where $f_{\alpha}(x)$ is the TFEC for the element
$\alpha$ (see appendix~\ref{append_part}). We can define the probabilities for the RNA molecule of being in the F and the UF
states by $p_{0}$ and $p_{1}$ respectively,
\begin{eqnarray}
p_{\sigma}(X_{T})=\frac{Z_{\sigma}(X_{T})}{Z(X_{T})}, \;\;{\rm with}\;\; \sigma=0,1~.
\label{r2}
\end{eqnarray}
The thermodynamic value of any observable ${\cal O}$ can be expressed in
terms of these probabilities,
\begin{eqnarray}
\langle {\cal O} \rangle=p_{0}\langle{\cal O} \rangle_{0}+p_{1}\langle {\cal O}\rangle_{1}~.
\label{r3}
\end{eqnarray}
At the midpoint of the transition both states are equally probable, 
\be
p_{0}(X_{T}^{c})=p_{1}(X_{T}^{c}) \;\;{\rm or}\;\;\;Z_{0}(X_{T}^{c})=Z_{1}(X_{T}^{c})~,
\label{w1}
\ee 
where these functions have been defined in \eqq{r1b}{r1c} and
(\ref{r2}). Hence the midpoint of the transition in the mixed-ensemble
is defined by the value of the control parameter $X_{T}^{c}$ that
verifies (\ref{w1}).
\subsection{Computation of the transition force $F^c$, the TFEC and the different 
contributions to the reversible work.}
\label{WT}
\begin{figure}[H]
\begin{center}
%trebsp1.eps
\includegraphics[width=7cm,clip]{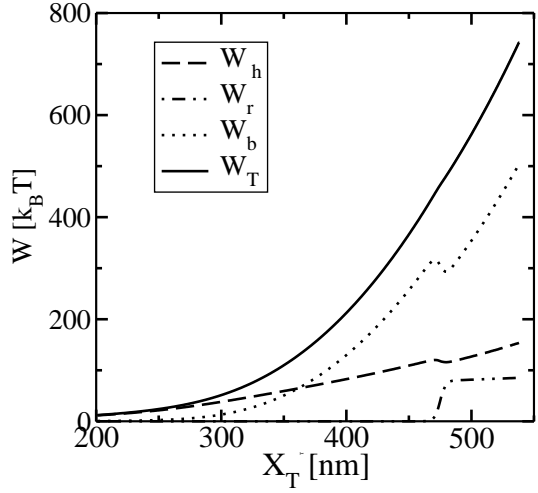}
\caption{{\small Different contributions to the reversible work
obtained from the partition function analysis: $W_{T}^{\rm rev}$,
$W_{h}^{\rm rev}$, $W_{b}^{\rm rev}$ and $W_{r}^{\rm rev}$ as a
function of $X_{T}$. Note that the smallest contribution to the total
work comes from the RNA molecule.}}
\label{f4}
\end{center}
\end{figure}

\begin{figure}[H]
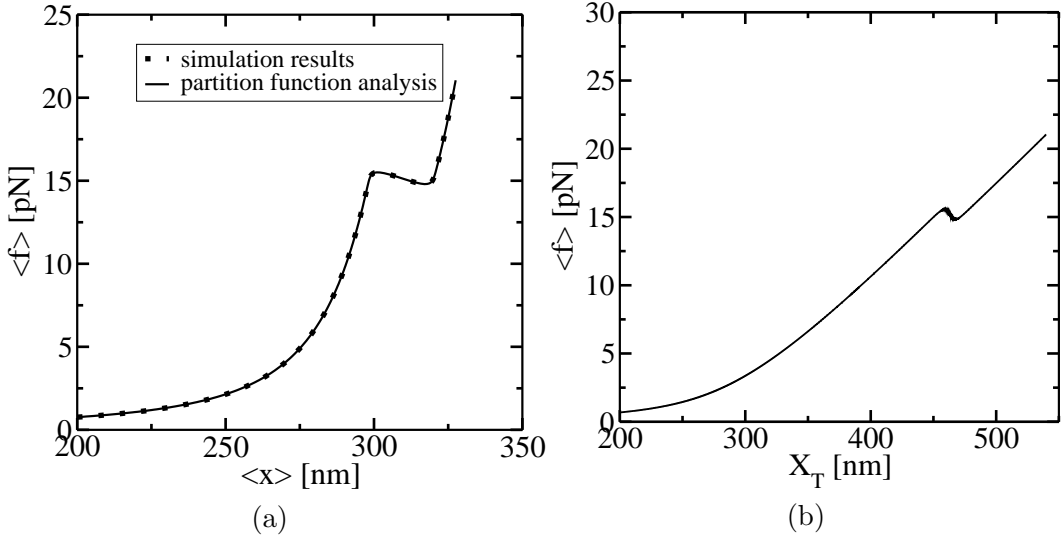

\begin{minipage}{7cm}
\begin{center}
%FEC1r.eps
\includegraphics[height=6.5cm]{FEcm.eps}\\
(a)
\end{center}
\end{minipage}
\begin{minipage}{7cm}
\begin{center}
%FEC1p.eps
\includegraphics[height=6.5cm]{fecxt.eps}\\
(b)
\end{center}
\end{minipage}
\caption{\small{ (a) The continuous line corresponds to the results
obtained from the numerical computation of the TFEC. It is also shown
the TFEC obtained by averaging over 1000 different trajectories in a
simulation of a pulling experiment as explained in
Sec.~\ref{simulation}. The pulling is carried out at an approximate
loading rate of $0.5\rm pN/s$ (see footnote~\ref{foot11}), slow enough
to generate a quasi-static process. One can observe that both curves
agree. (b) Mean force $\langle f\rangle$ as a function of the control
parameter $X_T$. Note that there is not an abrupt vertical drop of the
force at $X_T^c$. This is consequence of the narrow, yet
observable, region of coexistence around the midpoint of the
transition.}}
\label{f6}
\end{figure}

The force at the transition, $F^{c}$, is computed as the mean value of
 the force at $X_{T}^{c}$ given by (\ref{w1}). To reproduce
the experimental results obtained for the P5ab RNA molecule in 10mM $Mg^{2+}$
\cite{b10} we use the parameters
 given by Tables~\ref{table1} and ~\ref{table1b} getting $F^{c}=15.2\rm
 pN$~\footnote{\label{foot9}In our model we are considering the F state to be
 localized at $x_{r}=0$. However also the folded RNA has an end-to-end
 distance $d_r$ (the diameter of the hairpin) that tends to be aligned
 along the force axis when the force increases.  Modeling the F state
 as a rigid segment of length $d_r=2\rm nm$ one obtains $F^{c}=15.9\rm
 pN$ which is a bit farther from the experimental value.}. This value is
 close to the one reported from the experiments $F^{c}_{\rm
 exp}=14.5\pm 1\rm pN$ \cite{b10}. We also verify that the force at
 the transition $F^{c}$ is quite stable with respect to changes in the
 parameters of the problem used to model the handles and the bead
 trapped in the optical potential, such as the persistence and contour
 length of the handles, the spring constant and the bead
 radius. However, as the value of $F^{c}$ is highly influenced by the
 characteristics of the RNA molecule, we conclude that the dependence
 of the value of $F^{c}$ with the system is basically through the
 quantities $\Delta G^{0}$, $L_{r}$ and $P_{r}$.  
\begin{table}[H]
\begin{center}
\begin{tabular}{|c|c|c|c|c|}
\hline
$k_{B}T[\rm pNnm]$ &$k_{b}[\rm pN/nm]$ &$P_{h_{1}}=P_{h_{2}}[\rm nm]$  &$L_{h_{1}}=L_{h2}[n\rm m]$  &$E_{y}^{h_{1}}=E_{y}^{h_{2}}[\rm pN]$\\ 
\hline
4.14& 0.1 & 10 & 160& 1000\\ 
\hline
\end{tabular}
\end{center}
\caption{{\small{Summary table of the parameter values used to model
the handles and the bead in the optical trap. We use the value for the Young modulus corresponding to a dsDNA. The value for the other parameters have been taken
from \cite{b10}.}}\label{table1}}
\end{table}
\begin{table}[H]
\begin{center}
\begin{tabular}{|c|c|c|c|c|}
\hline
$P_{r}[\rm nm]$& $L_{r}[\rm nm]$ &$E_{y}^{r}[\rm pN]$ & $\Delta G^{0}[k_{B}T]$& $N \rm(number~ pair~ bases)$\\
\hline
1& 28.9 & 800 & 59 & 22\\ 
\hline  
\end{tabular}
\end{center}
\caption{{\small{Summary table of the parameter values used to model the RNA molecule. We use the value for the Young modulus corresponding to a ssDNA. The value for the other parameters have
been taken from \cite{b10}.}}\label{table1b}}
\end{table}
 Another
 interesting magnitude to measure is the reversible work $W_{T}^{\rm
 rev}$ done upon the system when pulling from an initial value of
 $X_T=X_T^0$ up to a final value of $X_T$. This work is given by 
\begin{eqnarray}
W_{T}^{\rm rev}(X_T)=G_{X_T}-G_{X_T^0}=\Delta G_{X_T}~,\rm{with}\nonumber \\
G_{X_T}=-k_{B}T\ln(Z(X_T))=-k_{B}T\ln(Z_0(X_T)+ Z_1(X_T) )~,
\label{w6}
\end{eqnarray}
where we used \eq{r1a}. The total reversible work in \eq{w6} defines
the change in the free energy of the system. On the other hand the reversible work exerted
upon the whole system is equal to the sum of reversible work exerted
on each element $W_{h}^{\rm rev},~W_{b}^{\rm rev},~ W_{r}^{\rm rev}$
(handles 1 and 2, bead and RNA molecule) by changing the total
end-to-end distance from the initial to the final value of $X_{T}$:
\begin{eqnarray}
W_{T}^{\rm rev}(X_T)=W_{b}^{\rm rev}(X_T)+W_{h}^{\rm rev}(X_T)+W_{r}^{\rm rev}(X_T)~,~~\rm {where}\label{w4a}\\
W_{b}^{\rm rev}(X_T)=\langle \Delta V_{b}\rangle=p_{0}\langle \Delta V_{b}\rangle_{0}+p_{1}\langle \Delta V_{b}\rangle_{1}~,\label{w4b}\\
W_{h}^{\rm rev}(X_T)=\langle W_{h} \rangle= \sum_{i=1}^{2}\Bigl[p_{0}\langle W_{h_{i}}\rangle_{0}+p_{1}\langle W_{h_{i}}\rangle_{1}\Bigr]~,\label{w4c}\\
W_{r}^{\rm rev}(X_T)=\langle W_{r} \rangle=p_{1}(\langle W_{r}\rangle_{1}+\Delta G^{0})~,
\label{w4d}
\end{eqnarray}
where we used (\ref{r3}). The functions $\Delta V_{b}$, $W_{h}$
and $W_{r}$ correspond to the change in the potential energy of the
bead in the optical trap and the work exerted upon both handles and
the RNA molecule by moving the total end-to-end distance from the
initial to the final value of $X_{T}$ respectively.  In Fig.~\ref{f4}
we show the different contributions to the total work $W_{h}^{\rm
rev}$ , $W_{b}^{\rm rev}$ and $W_{r}^{\rm rev}$ as a function of
$X_{T}$ as derived from the numerical computation of $Z(X_{T})$. We
also show the work $W_{T}^{\rm rev}$ exerted upon the whole
system. Both computations \eqq{w6}{w4a} overlap in a single curve as
expected.  Finally in Fig.~\ref{f6} (a) we represent the
TFEC for the subsystem $S$, $\langle f\rangle$\footnote{\label{foot10}At a given
state of the system (determined by a given value of $X_T$) all the
forces $f_{\alpha}$ are equal in average. Therefore the value of $\langle
f\rangle$ referring to the mean force acting on the bead $B_1$ (Fig.~\ref{f1})
coincides with the force $\langle f_{\alpha}\rangle$ acting on each element $\alpha$
as well as on the subsystem $S$.} versus $\langle x\rangle$. This is obtained by numerical
computation of the partition function using the relation, 
\be
\langle f\rangle =-\frac{\partial G_{X_{T}}(X_{T})}{\partial
X_{T}}=k_{B}T\frac{\partial \ln
Z(X_{T})}{\partial X_{T}}=k_{b}(X_{T}-\langle x\rangle).
\label{new1}
\ee
We also present the results obtained by averaging 1000 different trajectories
in a simulation of a pulling experiment as described later in
Sec.~\ref{simulation}, and both curves show very good agreement. In 
Fig.~\ref{f6} (b) we plot the mean force $\langle f\rangle$ as a function of
the control parameter $X_T$.

\subsection{Reversible work across the transition}
\label{Wrip}

\begin{figure}[H]
\begin{center}
\includegraphics[width=7cm,clip]{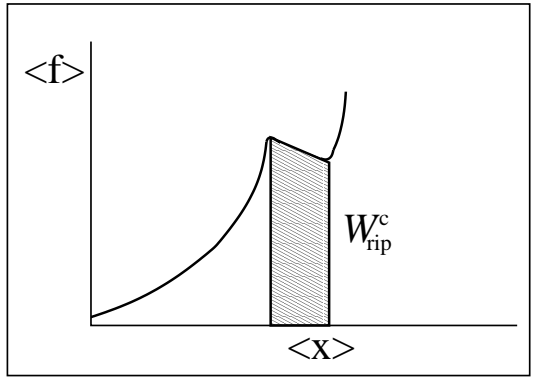}
\caption{\small{The shadow area under the TFEC along the transition 
corresponds to the quasi-static work $W_{\rm rip}^{c}$ (schematic representation).}}
\label{fWrip}
\end{center}
\end{figure}

 The quasi-static work $W_{\rm rip}^{c}$ exerted upon the subsystem $S$
across the transition is the area under the TFEC (Fig.~\ref{fWrip}),
 $\langle f\rangle
(\langle x\rangle )$, from $\langle x\rangle=\langle x^{c}\rangle_{0}$
to $\langle x\rangle=\langle x^{c}\rangle_{1}$, where the super-index $c$
indicates that the system is at the midpoint of the transition where
$X_{T}=X_{T}^{c}$ (\ref{w1}),
\begin{eqnarray}
W_{\rm rip}^{c}={\int}_{\langle x^{c}\rangle_{0}}^{\langle x^{c}\rangle_{1}}dy\langle f\rangle(y)= V_{b}(X_{T}^{c}-\langle x^{c}\rangle_{1})-V_{b}(X_{T}^{c}-\langle x^{c}\rangle_{0})~.
\label{rb4}
\end{eqnarray}

At the midpoint of the transition both states are equally populated and
(\ref{w1}) holds. Therefore identifying (\ref{r1b}) and (\ref{r1c}), we
can write (\ref{rb4}) as:
\begin{eqnarray}
W_{\rm rip}^{c}={{\Delta}G^{0}+W_{r}^{c}+{\Delta}W_{h}^{c}}~,
\label{r4}
\end{eqnarray}
where the functions with a super-index $c$ are evaluated at the mean
value of their variables at the critical extension $X_{T} ^{c}$. The
$W_{r}$ is the loss of entropy of the RNA molecule along the
transition due to the stretching and is given by (\ref{rr1}), and the
${\Delta}W_{h}$ is the change of free energy of the handles across the
transition computed as:
\begin{eqnarray}
{\Delta}W_{h}=W_{h_{1}}(\langle x_{h_{1}}\rangle_{1})+W_{h_{2}}(\langle x_{h_{2}}\rangle_{1})-W_{h_{1}}(\langle x_{h_{1}}\rangle_{0})-W_{h_{2}}(\langle x_{h_{2}}\rangle_{0})~.
\label{r5}
\end{eqnarray}
Eq.~(\ref{r4}) tells us that the quasi-static work $W_{\rm rip}^{c}$
coincides with the change of free energy of the different elements that
form the subsystem $S$ across the transition. This $W_{\rm rip}^{c}$ is
experimentally measurable as the area under the rip observed in the TFEC
corresponding to the F-UF transition (Fig.~\ref{fWrip}). Therefore
we can use (\ref{r4}) to estimate $\Delta G^{0}$ from the TFEC, as
 explained in the next section.

\subsection{Estimate of $\Delta G^{0}$ from the TFEC}
\label{compWrip}

\begin{figure}[H]
\begin{center}
\includegraphics[width=7cm,clip]{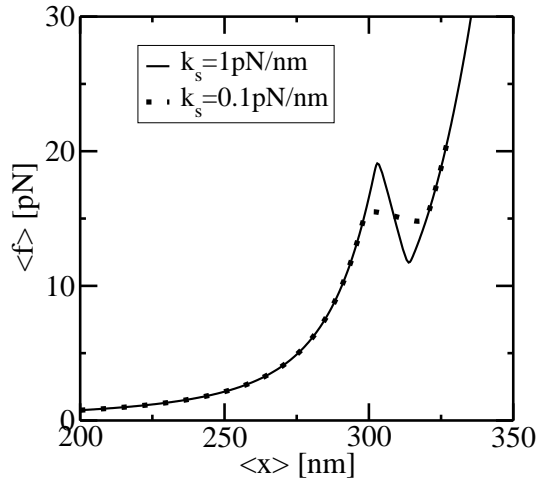}
\caption{\small{TFEC corresponding to two systems with handles and
RNA characterized by the parameters given in Tables~\ref{table1},~\ref{table1b} and with
an optical trap stiffness $k_{b}=0.1\rm pN/nm$ and $k_{b}=1\rm pN/nm$ respectively. Note that the slope at the
transition (rip) is proportional to $-k_{b}$.}}
\label{f10}
\end{center}
\end{figure}

\begin{figure}[H]
\begin{center}
\includegraphics[width=7cm,clip]{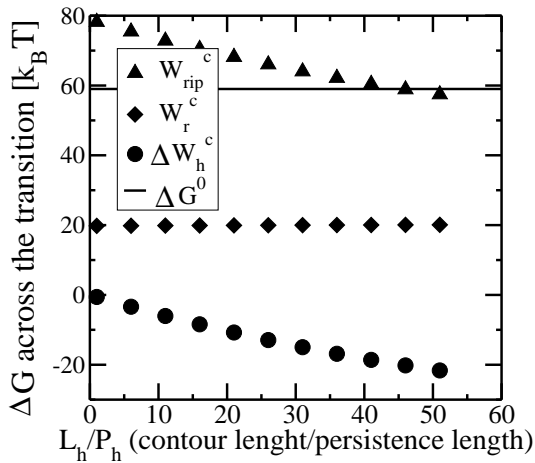}
\caption{\small{The different contributions to the free
energy change across the transition presented as a function of the ratio
$L_{h}/P_{h}$.}}
\label{f10b}
\end{center}
\end{figure}

In Fig.~\ref{f10} we show two TFECs obtained from the partition function analysis corresponding to two systems  with different $k_{b}$ but with the
same handles and RNA molecule with parameters given in Tables~\ref{table1} and \ref{table1b} respectively. We use (\ref{r4}) in
order to extract the value of $\Delta G^0$ by computing 
$W_{\rm rip}^{c}$ as the
area under the rip in the TFEC (Fig.~\ref{fWrip}).
 As expected for an harmonic trap (\ref{new1}), the TFEC in Fig.~\ref{f10} shows an slope at the transition (rip) proportional to $-k_{b}$.
To obtain the different contributions to (\ref{r4}) using the WLC model \cite{b25} we first estimate  
$\langle x_{r}\rangle_{1}$ and $\langle x_{h_{i}}\rangle_{1}$ as the RNA and the handle $i$ extension  at the force value after the rip, $\langle f\rangle_{1}$, respectively. In an analogous way we estimate the $\langle x_{h_{i}}\rangle_{0}$ as the handle $i$ extension at the force before the rip $\langle f\rangle_{0}$. Then we compute $W_{r}^{c}$ and $\Delta W_{h}^{c}$ given by \eq{rr1} and \eq{r5} respectively using the WLC model \cite{b25} for the TFEC $f_{\alpha}(x)$ for the element $\alpha$ with $\alpha=h_{1},h_{2},r$ (handles 1 and 2 and RNA respectively).
 Finally, we compute the area under the TFEC across the transition (rip) in order to obtain $W_{\rm rip}^{c}$ and use \eq{r4} to extract $\Delta G^{0}$. In Table \ref{tablew} we
show the results obtained.
\begin{table}[H]
\begin{center}
\begin{tabular}{|c|c|c|c|c|}
\hline
$k_{b}[\rm pN/nm]$ &$W_{r}^{c}[k_{B}T]$ &$\Delta W_{h}^{c}[k_{B}T]$ &$W_{rip}^{c}[k_{B}T]$ & $\Delta G^{0}[k_{B}T]$\\
\hline
0.1& 20 & -8.5 & 70.5 & 59\\ 
\hline
1& 17 & -41 & 35 & 59\\ 
\hline  
\end{tabular}
\caption{{\small{Different contributions to the free energy change across the transition. As expected the value of $\Delta G^{0}$ is independent of the other parameters of the system.}}\label{tablew}}
\end{center}
\end{table}
Note that the contribution $\Delta W_{h}^{c}$ is negative because when
the RNA molecule opens the force relaxes and the handles contract, hence
the free energy of the handles across the transition decreases.
Neglecting the contribution that comes from the handles across the
transition is a typical approximation often applied to experimental
results. However we note here that this is not always possible as this
contribution can be large. In the previous example, even in the case of
small $k_{b}$, we would loose $8k_{B}T$ in the balance equation
\eq{r4}. The best condition to apply this approximation is to use
handles characterized by a small ratio $L_{h}/P_{h}$ as compared to the
corresponding value for the RNA molecule ($L_{h}/P_{h}<<L_{r}/P_{r}$)
and a potential well with stiffness as small as possible (i.e, small
$k_{b}$). In Fig.~\ref{f10b} we show, for a small value of $k_{b}$
($k_{b}=0.1\rm pN/nm$), how the different contributions to \eq{r4} change
when considering systems with different values for the ratio
$L_{h}/P_{h}$. The stretching contribution to the UF state of the RNA,
$W_{r}^{c}$, does not change when modifying the magnitude $L_{h}/P_{h}$,
because the forces at which the transition occurs are quite stable under
changes of $L_{h}/P_{h}$. However, the magnitude of the contribution
$\Delta W_{h}^{c}$ tends to notably increase as $L_{h}/P_{h}$ becomes larger.

\section{Simulation of a pulling experiment}
\label{simulation}
As the RNA molecule unfolds-refolds in timescales much larger than the typical
relaxation time of handles and bead, we can consider an
instantaneous relaxation for these latter elements to solve the
dynamical equations. This hypothesis is valid as long as the data is 
collected at frequencies smaller than the relaxational frequency of the
bead, $10^{5}Hz$, that is the element with largest relaxation time
(see Sec.~\ref{parts}). The dynamics for the RNA molecule is governed 
 by the master equation for the probability $p_{\sigma}$ \eq{r2}, 
\begin{eqnarray} 
\frac{dp_{0}}{dt}=-k_{\rightarrow}p_{0}+ k_{\leftarrow}p_{1}~,\nonumber\\
\frac{dp_{1}}{dt}=-k_{\leftarrow}p_{1}+ k_{\rightarrow}p_{0}~,
\label{meq}
\end{eqnarray}
where $k_{\rightarrow}$ and $k_{\leftarrow}$ are the unfolding and folding
 rates respectively. To simulate a pulling experiment we parallelly solve numerically the
 partition function of the system finding the mean extension and force 
for each element and do a numerical simulation of the dynamical model for 
the RNA \eq{meq}. We implement the following algorithm:
\begin{itemize}
\item We increase $X_{T}$ by $v\Delta t$, where $v$ is the pulling
speed, i.e the velocity at which the micropippete is pulled,
$v=\dot{X_{T}}$, and $\Delta t$ is the iteration time, hence
$\frac{1}{\Delta t}$ is the frequency at which data is
collected~\footnote{\label{foot11} The relation between the pulling
speed $v$ and the loading rate $r$ (velocity at which the force
increases) can be found using the relation between the force and
displacement increments, ${\Delta}f=k_{\rm eff}(f){\Delta}X_{T}$, as \be
r=vk_{\rm eff}~,
\label{rv}
\ee
where $k_{\rm eff}$ is the effective stiffness of the system, computed
as:

\be
k_{\rm eff}=\frac{d\langle f(X_{T})\rangle}{dX_{T}}=[\frac{1}{k_{b}}+\frac{1}{k_{x}}]^{-1},
\label{s0}
\ee
where $k_{x}$ has been defined in (\ref{kx}) and $k_{b}$ is the
stiffness of the optical trap. The F-UF transition for a small single
domain of RNA typically occurs at forces in the range $8-20\rm pN$. At these
forces the system verifies that $k_{b}$ is much smaller than the
stiffness of the handles and the RNA molecule, $k_{h_{1}},k_{h_{2}}$ and
$k_{r}$, therefore we can safely assume $v=r/k_{b}$.}.
\item We compute the new $\langle f\rangle$ and $\langle x\rangle$
iteratively using the saddle point equations for the partition
function. To these mean values we add Gaussian fluctuations of zero
mean and variance given by (\ref{flumix}). We then obtain the FEC,
$f(x)$, that should qualitatively reproduce the experimental one.

\item The RNA molecule is then unfolded (if it is in the folded state
$\sigma=0$) or folded (if it is in the unfolded state
$\sigma=1$) with a probability
$k_{\rightarrow}(X_{T})\Delta t$ and $k_{\leftarrow}(X_{T})\Delta t$ respectively,
where $\Delta t$ is the iteration time. These probabilities come from the discretization of the master equations \eq{meq}.  
The unfolding and folding rates,
$k_{\rightarrow}$ and $k_{\leftarrow}$, correspond to the rates for an
activated process characterized by a barrier $B(X_{T})$ and a difference
of free energy between the F and UF states $\Delta G(X_{T})$ (Fig.~\ref{f3} (a)),
\begin{eqnarray}
k_{\rightarrow}(X_{T})=k_{0}\exp[-\beta B(X_{T})]\nonumber\\
k_{\leftarrow}(X_{T})=k_{0}\exp[\beta(-B(X_{T})+\Delta G(X_{T}))]~,
\label{ss1}
\end{eqnarray}
where $k_{0}$ is an attempt frequency. These rates satisfy the detailed
balance condition,
\be
\frac{k_{\rightarrow}(X_{T})}{k_{\leftarrow}(X_{T})}=\exp[-\beta\Delta G(X_{T})]~.
\label{ss1b}
\ee
The expressions of $\Delta G(X_{T})$ and $B(X_{T})$ are derived in appendix ~\ref{append_foldunfold} using the partition function analysis.    
\end{itemize}

In the simulations presented in Secs.~\ref{fec},~\ref{fraction} we use the
parameters given in Tables \ref{table1} and \ref{table1b}. In Table~\ref{table2} we show the
values of the kinetic parameters we use, such as the rate of unfolding at
zero force $k_{0}\exp(-\beta B^{0})$ and the number of opened bases
$n^{*}$ that characterizes the location of the transition state (see
Sec.~\ref{twostates})\footnote{\label{foot11b} The value of $n^{*}$ determines 
the distance from the barrier to the folded conformation, $x_{1}(X_{T})$. 
With the assumption that  $n^{*}$ does not depend on $X_{T}$ 
(see footnote \ref{foot5}) one can derive $x_{1}(X_{T})$ using the WLC model
 (\ref{p2}) with $P=P_{RNA}$ and $L_{o}=L_{RNA}\frac{n^{*}}{N}$,
$x_{1}(X_{T})=x(f)$ where $f$ is the mean force acting upon the system 
when the RNA molecule is in the transition state.}.
\begin{table}[H] 
\begin{center}
\begin{tabular}{|c|c|c|c|c|}
\hline
$k_{0}\exp(-\beta B^{0})$ &$n^{*}$\\
\hline
$e^{-30}\approx 10^{-13}$& 12 \\ 
\hline  
\end{tabular}
\caption{{\small{Parameters used to characterize the kinetics of
folding-unfolding of RNA. They are chosen in order to reproduce the
experimental kinetics results obtained with the hairpin P5ab
\cite{b10}.}}\label{table2}}
\end{center}
\end{table}
 
In what follows we present the results of our simulations performed 
to analyze the following
aspects: i) Obtaining FECs in the mixed ensemble; ii) Computation of
the fraction of forward (reverse) trajectories that have at least one
refolding.

\subsection{Force-extension curve results (FEC)}
\label{fec}
\begin{figure}[H]
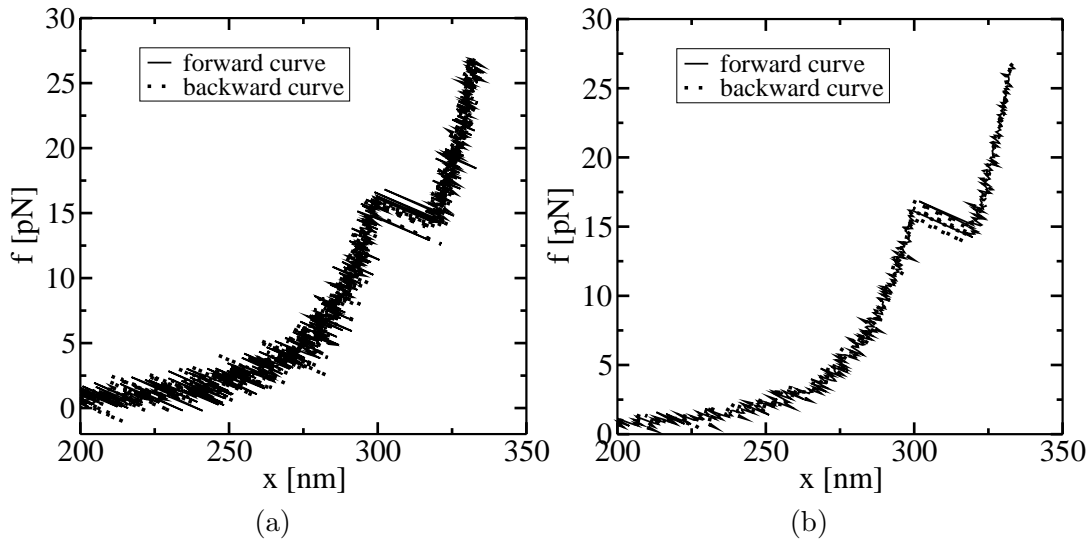

\begin{minipage}{7cm}
\begin{center}
%FEC1r.eps
\includegraphics[height=6.5cm]{FEC1r1.eps}\\
(a)
\end{center}
\end{minipage}
\begin{minipage}{7cm}
\begin{center}
%FEC1p.eps
\includegraphics[height=6.5cm]{FEC1pp.eps}\\
(b)
\end{center}
\end{minipage}
\caption{\small{Results for the FEC obtained from the simulation of a pulling experiment with $r=1 \rm
pN/s$. The iteration time used in the simulation is $\Delta t=10^{-2}\rm
s$. In (a) we show the results of calculations at each time step. In (b) at
we present their average over five consecutive time steps.}}
\label{f7}
\end{figure}

\begin{figure}[H]
\begin{center}
%FEC10.eps
\includegraphics[width=7cm,clip]{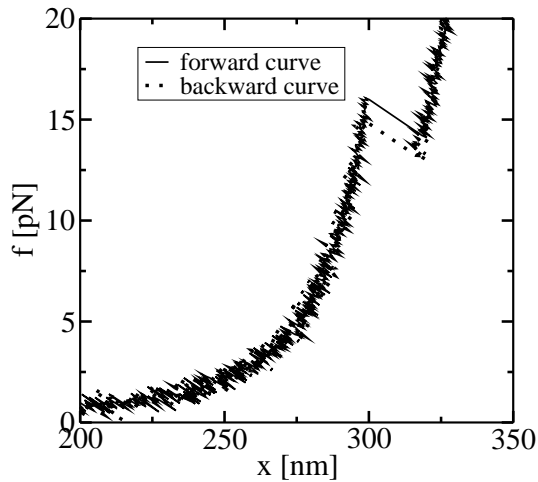}
\caption{\small{ Results for the FEC obtained from the simulation of a pulling experiment for a $r=50\rm pN/s$. The iteration time used in the simulation is $\Delta t=10^{-2}\rm s$. At this pulling speed ($v$, see footnote \ref{foot11}) the process is not in equilibrium and hysteresis is observed around the transition.}}
\label{f9}
\end{center}
\end{figure}

In Figs.~\ref{f7} and \ref{f9} we show the resulting FEC of our
simulations for the values used in the experiment of Liphardt et al.
\cite{b10} shown in Tables~\ref{table1}, \ref{table1b} and \ref{table2}
corresponding to a P5ab RNA molecule and for a loading rate of $r=1\rm pN/s$ and of $r=50\rm pN/s$ respectively. We do the simulation for the forward and
reverse processes where $X_T$ increases and decreases in time
respectively.
\begin{figure}[H]
\begin{center}
\includegraphics[width=7cm,clip]{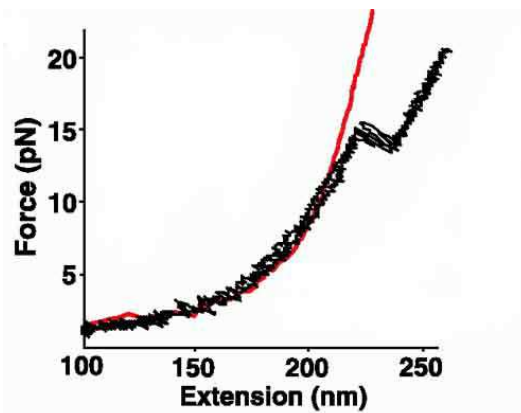}
\caption{\small{ Experimental FEC for p5ab obtained in \cite{b10}. The continuous line corresponds to the WLC curve for the handles.}}
\label{f8b}
\end{center}
\end{figure}
As shown in Fig.~\ref{f7}, at a loading rate of $1\rm pN/s$ different transition jumps
are observed along both the forward and reverse processes, 
because the pulling speed ($v$, see footnote \ref{foot11}) is slow enough.
 In Fig.~\ref{f7} (a) we represent the FEC
resulting from the computed value  of $f$ and $x$ at each iteration
whereas in Fig.~\ref{f7} (b) we show the FEC obtained after averaging the results over 
five consecutive iterations. The amplitude of the fluctuations observed
in Fig.~\ref{f7} (b) notably decreases. These values
appear compatible with those found in the experimental data. Comparing
these simulations results with the experimental FEC \cite{b10} shown
in Fig.~\ref{f8b} we find a qualitative agreement, the shape of the
curve around the transition region is qualitatively
reproduced. 
 However, we find some discrepancies: (i) The simulated curve is shifted in the $x$ direction in comparison with the experimental one. This is because experimentally the quantity measured is not the absolute value of the distance 
$x$ but its relative changes. Therefore
in Fig.\ref{f8b} the extension represented in the $x$-axis corresponds to
changes in the value of $x$ with respect to an initial extension of approximately  $100\rm nm$. (ii) As the force increases the experimental curve separates from the theoretical WLC prediction and therefore from the simulated results. The agreement  can be improved by considering bigger values for the Young modulus of the handles and of the ssRNA. Furthermore, extending the RNA molecule model to include intermediate configurations, which depend on the number of opened bases $n$, one realizes that the cooperative transition might not be between the F ($n=0$) and UF ($n=N$) states, but between a partially folded and a partially unfolded states. For instance, for the P5ab RNA molecule the cooperative folding-unfolding transition is between the state $n=3$ and the state $n=N$ ~\cite{CocMonMar}. This means that typically the first 3 base pairs open before the 
transition occurs, increasing the extension of the handles. 
   
Fig.~\ref{f9} shows the FEC corresponding to a pulling
process carried out at a loading rate of $r=50\rm pN/s$. At this pulling speed the
process is not in equilibrium and hysteresis effects are observed around the
transition region.

\subsection{Fraction of trajectories that have at least one refolding}
\label{fraction}
\begin{figure}[H]
\begin{center}
%FEC10.eps
\includegraphics[width=7cm,clip]{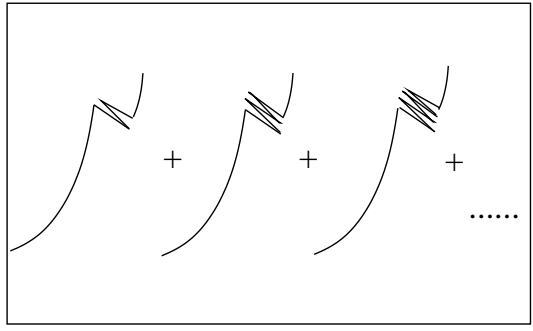}
\caption{\small{ Different trajectories that have at least one refolding. The ratio between this sum and the total number of trajectories gives the fraction $N_{F}$ ($N_{R}$) for the forward (reverse) process.}}
\label{f9b}
\end{center}
\end{figure}

\begin{figure}[H]
\begin{center}
%FEC10.eps
\includegraphics[width=7cm,clip]{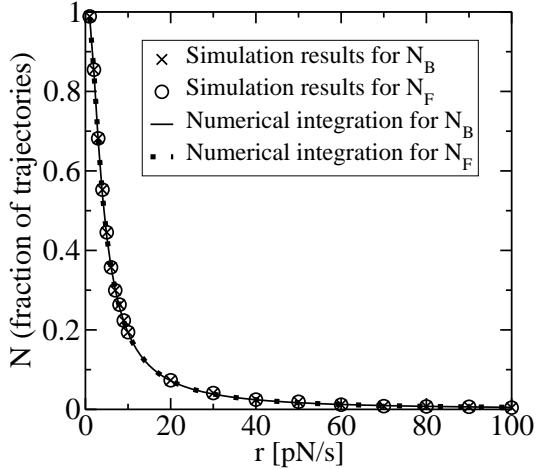}
\caption{\small{ The fraction $N_{F}$ and $N_{R}$ as a function of
$r$. It is shown the results obtained from a 5000 realizations of the
simulation of a pulling experiment and also the numerical integration of
(\ref{s10a}) or (\ref{s10b}) using the rates given by (\ref{s111}) characterized by the
following effective parameters:  $\tilde{B}\ln k_{0}=35.2k_{B}T$,
$\tilde{\Delta G}=70.4k_{B}T$, $\tilde{x}_{1}=9.75 \rm nm$ and
$\tilde{x}_{2}=9.35 \rm nm$. }}
\label{f11}
\end{center}
\end{figure}

We consider a system with a control parameter (generally denoted by $y$)
 that is pulled by
changing $y$ at certain speed $v(y)=\frac{dy}{dt}$. The forward
(reverse) pulling process starts at a initial value of the control
parameter $y_{i}$ ($y_{f}$) where the RNA is in the F (UF) state and
finishes at a final value of the control parameter $y_{f}$ ($y_{i}$)
where the RNA is in the UF (F) state. We then define $N_{F}$ and
$N_{R}$ as the fractions of forward and reverse trajectories that have
at least one refolding respectively (Fig.~\ref{f9b}).
These fractions are given by
\begin{eqnarray}
N_{F}=\int_{y_{i}}^{y_{f}}\int_{y}^{y_{f}}\frac{d\rho_{0}^{F}(y_{i},y)}{dy}\frac{d\rho_{1}^{F}((y,y')}{dy'}dy'dy~,\label{s10a}\\
N_{R}=\int_{y_{f}}^{y_{i}}\int_{y}^{y_{i}}\frac{d\rho_{1}^{R}(y_{f},y)}{dy}\frac{d\rho_{0}^{R}(y,y')}{dy'}dy'dy~,
\label{s10b}
\end{eqnarray}
where the first integral in the right-hand side of both equations
accounts for the probability of unfolding (folding) before a certain
value of the control parameter $y$ is reached and the second integral
accounts for the probability of refolding once the RNA molecule has been
unfolded (folded). The function $\rho_{\sigma}^{F(R)}(z,z')$ is the
probability for the RNA molecule of remaining at state $\sigma$ until
$y=z'$ starting at $y=z$ in the forward (reverse) process.
The $\rho_{\sigma}$ is solution of the master equation
\be
\frac{d\rho_{0}^{F(R)}(y,y')}{dt}=-k_{\rightarrow}(y')\rho_{0}^{F(R)}(y,y')~;
\label{m5}
\ee
\be
\frac{d\rho_{1}^{F(R)}(y,y')}{dt}=-k_{\leftarrow}(y')\rho_{1}^{F(R)}(y,y')~,
\label{m5b}
\ee with initial condition $\rho_{\sigma}^{F(R)}(y,y)=1$.  In
appendix~\ref{NFR} we prove that the fraction $N_{F}$ is equal to
$N_{R}$ if the perturbation protocol for the control parameter is
symmetric, i.e. if the velocities along the forward and reverse
process verify $v_{F}(y)=-v_{R}(y)$.  In our analysis the control
parameter $y$ corresponds to the total distance $X_T$ and the
folding-unfolding rates are given in \eq{ss1}. The detailed analytical
expressions have been given (\ref{rates1a},\ref{rates1b}) in the
appendix~\ref{append_foldunfold}. However, working with these rates in
order to do analytical computations appears quite cumbersome and it is
preferable to simplify them. For analytical purposes we will consider
effective rates where the functions $B^{1}$, $\Delta G^{1}$ given by
(\ref{rates1c}) and $x_{1}$ and $x_{2}$ (the distances from the F and
UF states to the transition state along the $x$-axis, see
Fig.~\ref{f3}) are effective parameters independent of $X_T$, that we
call $\tilde{B}$, $\tilde{\Delta G}$, $\tilde{x}_{1}$ and
$\tilde{x}_{2}$, obtaining
\begin{eqnarray}
k_{\rightarrow}(f_{0})&=&k_{0}\exp[\beta(-\tilde{B}+f_{0}\tilde{x}_{1}-\frac{1}{2}k_{b}\tilde{x}_{1}^{2})]~,\nonumber\\
k_{\leftarrow}(f_{1})&=&k_{0}\exp[\beta(-\tilde{B}-f_{1}\tilde{x}_{2}+\tilde{\Delta G}-\frac{1}{2}k_{b}\tilde{x}_{2}^{2})]~,
\label{s111}
\end{eqnarray}
where the force $f_{\sigma}$ ($\sigma=0,1$) corresponds to the force
acting upon the system at a given value of $X_T$ when the RNA is in
the state $\sigma$~\footnote{\label{foot12}The approximation \eq{s111}
where force does not fluctuate near the transition is well
justified. In fact, when the RNA is in a given state (folded or
unfolded) the magnitude of force fluctuations is negligible (the r.m.s
is in the range $10^{-3}-10^{-2}\rm{pN}^{2}$), so one can consider the
instantaneous force equal to the mean force.}. In what follows we will call the
dynamics generated by the effective rates \eq{s111} the effective dynamics
and the ones generated by the rates (\ref{rates1a},\ref{rates1b})
the non-effective dynamics. The effective rates are an excellent approximation to the non-effective ones in the experimental regime (see appendix~\ref{app_singledomain}). The relation between the forces $f_{0}$ and
$f_{1}$ (for a fixed value of $X_T$) in \eq{s111} is given by
\be
f_{1}=f_{0}-k_{b}\tilde{x}_{m}~,
\label{f0f1}
\ee
 where $\tilde{x}_{m}$ is the distance between the F and UF states,
$\tilde{x}_{m}=\tilde{x}_{1}+\tilde{x}_{2}$. Using (\ref{f0f1}) it is
straightforward to see that the effective rates (\ref{s111}) satisfy
the detailed balance condition \eq{ss1b}.  We can now compute the
fractions \eqq{s10a}{s10b} as a function of the loading rate $r$.  In
Fig.~\ref{f11} we show the results obtained for the fractions $N_{F}$
and $N_{R}$ from the numerical computation of (\ref{s10a},\ref{s10b})
using the effective dynamics (\ref{s111}). We also show the results
obtained from the simulations for the fractions $N_{F}$ and $N_{R}$ as
a function of the loading rate $r$ and they agree pretty well.

Through the simulation we are able to compute the mean work exerted
upon the system as a function of $r$:
\begin{eqnarray}
\langle W(r)\rangle=\langle\sum_{i=1}^{n}f_{i}\Delta X_{T}\rangle~, 
\end{eqnarray}
where $\Delta X_{T}$ is the increase in the total end-to-end distance in
each iteration and $n$ is the total number of
iterations. The average is over different realizations of the simulation
of the pulling process. The total work is the sum of the reversible work
(i.e. the work measured in a quasi-static process for $r$ going to
zero), and the mean dissipated work that is also a function of $r$:
$\langle W(r)\rangle=W^{T}_{\rm rev}+\langle W_{\rm dis}(r) \rangle$.

\begin{figure}[H]
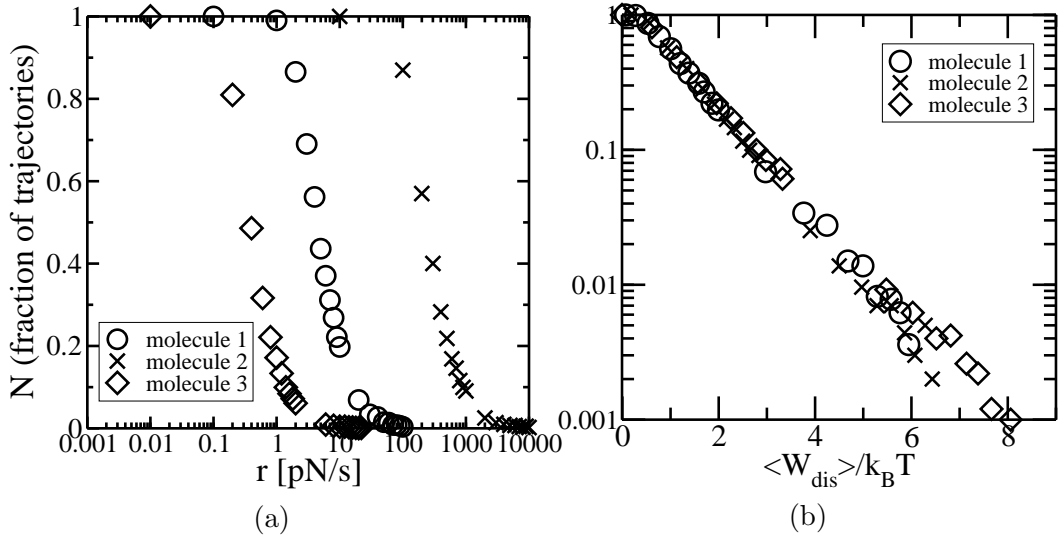

\begin{minipage}{7cm}
\begin{center}
%FEC1r.eps
\includegraphics[height=6.5cm]{N3r.eps}\\
(a)
\end{center}
\end{minipage}
\begin{minipage}{7cm}
\begin{center}
%FEC1p.eps
\includegraphics[height=6.5cm]{N3log.eps}\\
(b)
\end{center}
\end{minipage}
\caption{\small{(a) The fraction $N_{F}$ as a function of $r$ for
three different RNA molecules characterized by: Molecule (1) $\Delta
G^{0}=59k_{B}T$, $L_{r}=28.9\rm nm$, $N=24$, $n^{*}=12$ and $B^{0}\ln
k_{0}=29k_{B}T$. Molecule (2) $\Delta G^{0}=89k_{B}T$, $L_{r}=40\rm
nm$, $N=34$, $n^{*}=15$, $B^{0}\ln k_{0}=45k_{B}T$. Molecule (3)
$\Delta G^{0}=39k_{B}T$, $L_{r}=16.5\rm nm$, $N=14$, $n^{*}=9$ and
$B^{0}\ln k_{0}=19k_{B}T$. (b) The fraction $N_{F}$ as a function of
$\langle W_{\rm dis} \rangle$ in logarithmic scale for the three RNA
molecules considered in the left panel (a).}}
\label{f11b}
\end{figure}

\begin{figure}[H]
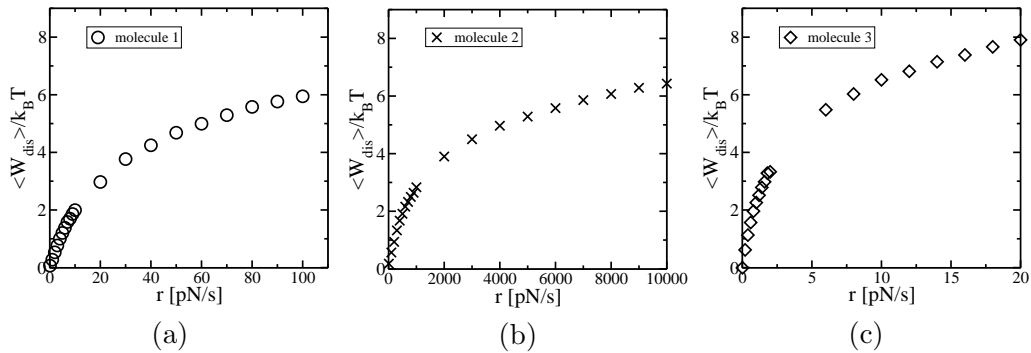

\begin{minipage}{4.5cm}
\begin{center}
%FEC1r.eps
\includegraphics[height=4cm]{Wdisr1.eps}\\
(a)
\end{center}
\end{minipage}
\begin{minipage}{4.5cm}
\begin{center}
%FEC1p.eps
\includegraphics[height=4cm]{Wdisr2.eps}\\
(b)
\end{center}
\end{minipage}
\begin{minipage}{4.5cm}
\begin{center}
%FEC1p.eps
\includegraphics[height=4cm]{Wdisr3.eps}\\
(c)
\end{center}
\end{minipage}
\caption{\small{ Mean dissipated work as a function of the loading rate $r$ for Molecule 1 in (a), for Molecule 2 in (b) and for Molecule 3 in (c). The characteristics for the three molecules are given in Fig.~\ref{f11b}. Note that the regimes studied are far from the linear response regime as the curves deviate from the straight lines. The deviation from the linear response regime arises at the range of $r$ where the fraction $N$ approaches to zero (Fig.~\ref{f11b} (a)).}} 
\label{f11c}
\end{figure}

We then consider the fraction $N_{F}$ for three different RNA
molecules characterized by different parameters $\Delta G^{0}$,
$L_{r}$, $N$ (total number of pair bases),  $n^{*}$ and $B^{0}\ln k_{0}$ and the results as a
function of $r$ are shown in Fig.~\ref{f11b} (a). When
we plot these fractions $N_{F}$ as a function of the mean dissipated
work $\langle W_{\rm dis} \rangle$ exerted upon the system we see that
the three curves corresponding to the three RNA molecules collapse to
a single curve as it is shown in Fig.~\ref{f11b} (b)
. This suggests that there is a generic dependence for the fraction $N_{F}$
as a function of $\langle W_{\rm dis} \rangle$. This dependence is not
surprising as the average dissipated work has been already
shown~\cite{b16} to be a useful quantity to characterize the
non-equilibrium regime. In particular, in the linear response regime,
the average dissipated work depends linearly on the loading rate $r$,
the proportionality constant being a function of the relaxation time
of the molecule, the unfolding free energy and the transition
force~\cite{b16}.
The collapse of all curves in Fig.~\ref{f11b} in a single curve is,
however, not restricted to the linear response regime. Indeed, we have
verified that in the regime $2k_BT<\langle W_{\rm dis} \rangle< 5
k_BT$, where deviations from the linear response regime are
observable (Fig.~\ref{f11c}), there is still a good collapse in
Fig.~\ref{f11b} (b) of the curves corresponding to the three molecules.
 Note that by measuring the fraction $N_{F}$ we can obtain information about
the value $\langle W_{\rm dis} \rangle$, and knowing the total work we
can extract the reversible work exerted upon the system. This provides
an alternative way to derive equilibrium information from
non-equilibrium experiments~\footnote{\label{foot12b} Several methods has been proposed and tested \cite{Jarzynski, Ritort}}.


\section{Unfolding of domains stabilized by $Mg^{2+}$ tertiary contacts}
\label{domains}

In the presence of magnesium ions ($Mg^{2+}$) the kinetics of the
unfolding process can change dramatically if tertiary contacts are
formed. In the  experiments done in
\cite{b10} two different RNA molecules were studied, P5ab and P5abc,
with and without $Mg^{2+}$. The results obtained in \cite{b10} show
that in presence of $Mg^{2+}$ there are two different situations:
\begin{itemize}
\item
If there is no formation of tertiary contacts the folding-unfolding
behavior does not change qualitatively. This might be consequence of
the electrostatic stabilization that acts homogeneously along the 
molecule; all base-pair hydrogen-bonds become more stable and the free
energy landscape changes in a homogeneous way. This induces a
slight increase of $\Delta G^{0}$ and $B^{0}$, resulting in a value of
$F^{c}$ that is a bit larger and a kinetics that is slower than in
absence of $Mg^{2+}$. Indeed, this is what seems to happen in the case
of the P5ab RNA molecule \cite{b10}.
\item
When tertiary contacts stabilized by $Mg^{2+}$ are formed the free
energy landscape changes drastically, in particular in the vicinity  
of the bases that
are involved in the formation of such tertiary contacts. Therefore the
kinetics slows down dramatically and the unfolding-folding process
changes totally, as observed with P5abc RNA \cite{b10}.
\end{itemize}   
In this Section we will focus on the study of molecules that form
tertiary contacts induced by $Mg^{2+}$. Experiments on the unfolding
kinetics of domains stabilized by $Mg^{2+}$ tertiary contacts show how
intermediate states are characterized by big barriers that are located
close to the folded state along the $x$-axis~\footnote{\label{foot12b}
Recent studies \cite{bb26} show that the domains stabilized by
$Mg^{2+}$ tertiary contacts are better characterized by kinetic models
with more than one barrier. However here we will consider the simpler
case of a single barrier per domain.}  \cite{b10,b11}, $x_{1}\ll
x_{m}$ (Fig.~\ref{f3} (a)). Consequently the height of the barrier $B$
is quite insensitive to the force (or $X_{T}$), meaning that when the
force exerted upon the system increases, $B$ decreases much slower
than the difference of free energy between both states, $\Delta G$.
Therefore big barriers and small values of $x_{1}$ imply slow
unfolding processes. In complex RNA molecules the domains stabilized
by the presence of $Mg^{2+}$-tertiary contacts are rate-limiting for
the unfolding of the whole molecule \cite{b27,b28,b29,b30}.  In these
conditions, even at very low loading rates, the probability of
refolding, once the domain is unfolded, is almost zero. The unfolding
of RNA molecules with $Mg^{2+}$ dependent barriers at experimental
loading rates ($r\approx3-5 {\rm pN/s}$) becomes a 'stick-slip'
process \cite{b11}. Therefore we can use the following transition
rates
\footnote{\label{foot13}Note that these rates do not verify the
detailed balance condition.}:
\be
k_{\rightarrow}(X_{T})=k_{o}e^{-B(X_{T})/k_{B}T}~, ~~k_{\leftarrow}(X_{T})=0~.
\label{m3}
\ee
These rates have been considered by Evans and Richie in the study of
bond failure \cite{b12,b13}. However, in order to get a realistic
modelization using the rates \eq{m3}, the system requires that at the
breakage force $f^{*}(X_T)$ (the force at which the molecule opens)
the UF state is more stable than the F state, or $\Delta
G(f^{*}(X_{T}))<0$. The breakage force changes from experiment to
experiment due to the stochastic nature of the unfolding
process. Therefore we expect that the distribution of breakage forces
goes to zero when approaching $f_{m}$, where $f_{m}$ verifies $\Delta
G(f_{m})=0$, i.e. the value of the force when the RNA is in the F state  at the midpoint of the
transition \eq{w1}, i.e $f_{m}=f_{1}(X_{T}^{c})$.  For such process, the kind of
information that one can get from the analysis of the distribution of
breakage forces $f^{*}$ is about the kinetics rather than the
thermodynamics.

In all the previous analysis we have considered the
study of single domain RNA molecules. Now we want to analyze molecules
that have more than one domain. In order to do that we extend the model
developed in preceding sections to describe more complex RNA molecules. Here
we consider how to extract kinetic information by analyzing the
breakage force distribution for the case of a multidomain RNA molecule
with a sequential unfolding of its domains. To this end, it is convenient
to analyze first the case of a single RNA domain stabilized by
$Mg^{2+}$ tertiary contacts. As this problem has been already
considered by several authors we collect some of the main results in
the appendix~\ref{app_singledomain}.

\subsection{Domains with $Mg^{2+}$- dependent barriers that unfold sequentially under a loading rate}
\label{sequential_unfolding}
\begin{figure}[H]
\begin{minipage}{6cm}
\begin{center}
%FEC1r.eps
\includegraphics[height=5.5cm]{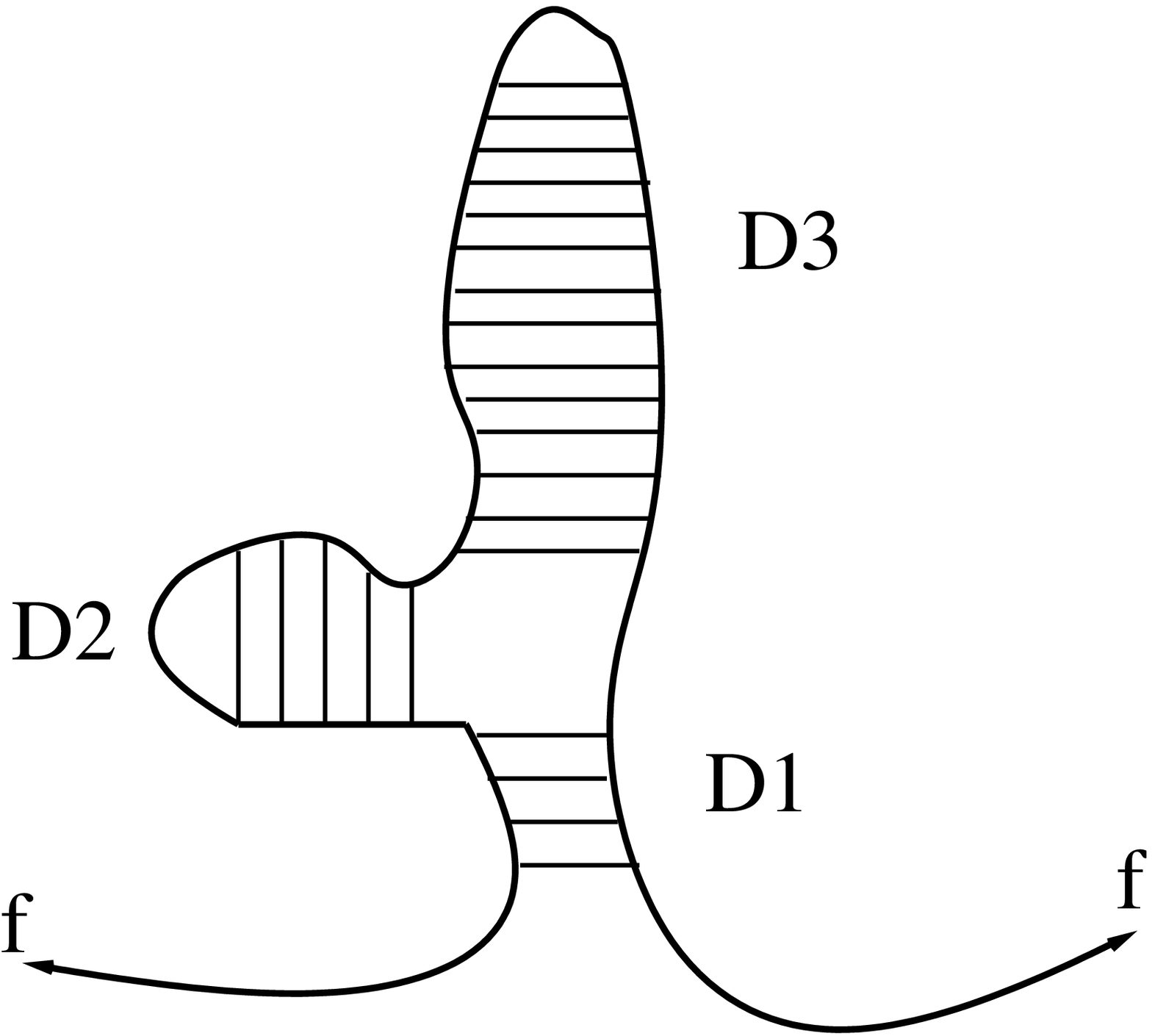}\\
(a)
\end{center}
\end{minipage}
\begin{minipage}{5.5cm}
\begin{center}
%FEC1p.eps
\includegraphics[height=4cm]{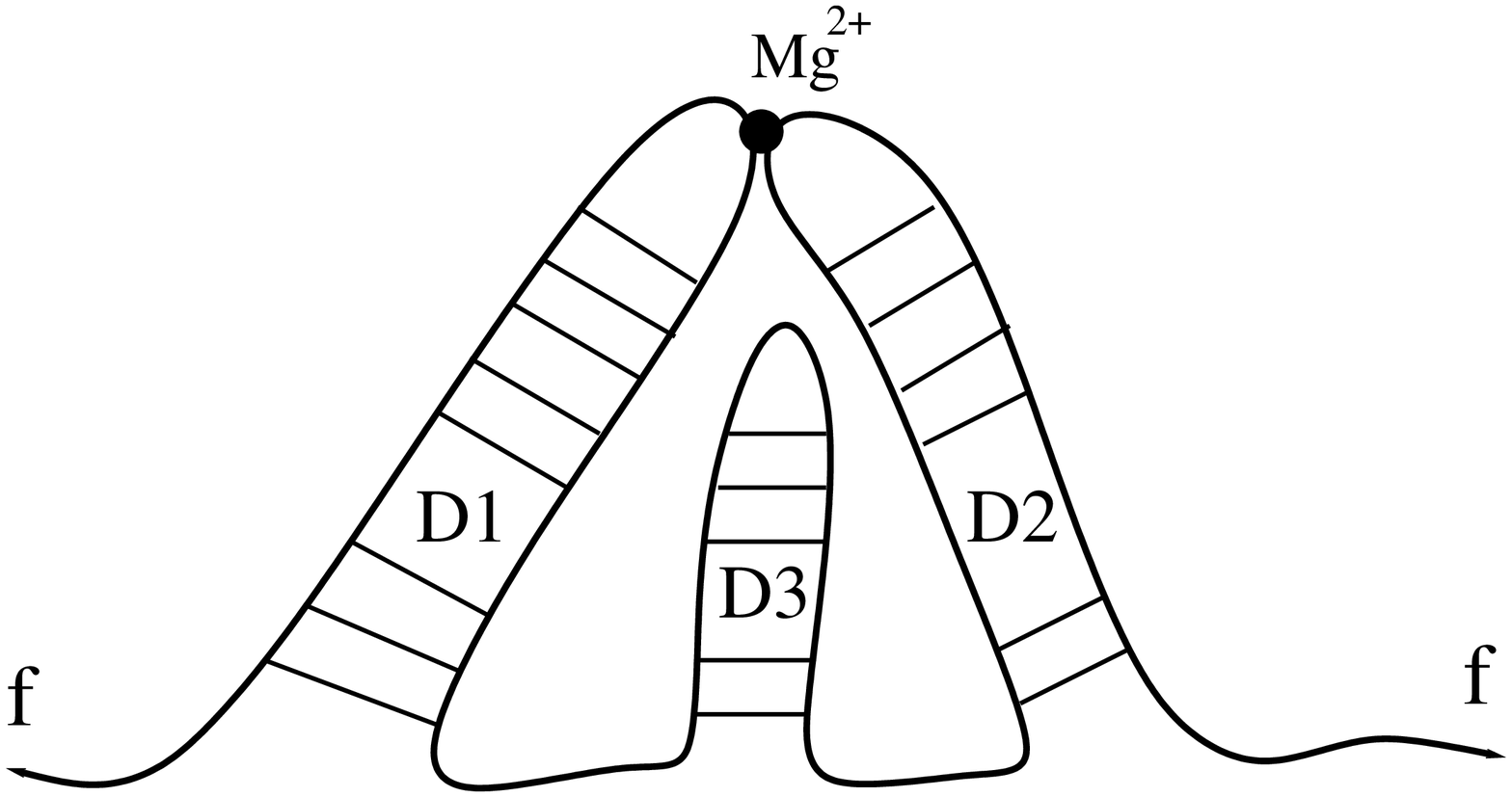}\\
(b)
\end{center}
\end{minipage}
\caption{\small{(a): Blockade of the force for certain domains
due to the connectivity of the molecule. The force can not act over
the domains D2 and D3 until D1 is not opened. (b): Blockade of
the force for certain domains due to the $Mg^{2+}$ tertiary
contacts. The domain D3 does not feel the force until the $Mg^{2+}$
tertiary contact breaks.}}
\label{f14}
%\end{center}
\end{figure}

In this section we want to investigate the applicability of 
 the model developed for a single
domain RNA to more complex RNA molecules such as a multidomain RNA molecule
with a sequential unfolding of its domains.  We consider a molecule
composed by different domains under the effect of an external force,
focusing on the case where the opening of the domains occurs in a
given sequential order. There are two situations that favor
a sequential unfolding of the domains.  The first one relies on the
topological connectivity of the molecule, that does not allow certain
domains to unfold before others have not yet opened (Fig.~\ref{f14} (a)). The second one is the blockade of the force 
induced by the most external tertiary contacts on the interior domains 
(Fig.~\ref{f14} (b)).

\begin{figure}[H]
\begin{center}
\includegraphics[width=7cm,clip]{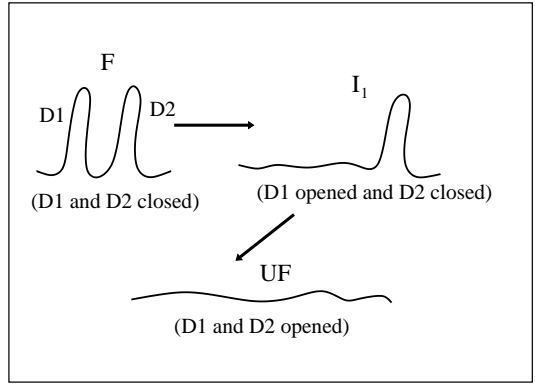}
\caption{\small{ Representation of the different states for a 2-domain
model corresponding to a molecule with two domains that sequentially
unfolds. The kinetics parameters of each domain are
$\tilde{x}_{1}^{(i)}$, $\tilde{x}_{m}^{(i)}$ and $\tilde{B}^{(i)}$,
where the super-index $i=1,2$ refers to the index of the domain. The
unfolding rate and the breakage force for the domain $i$ are denoted
as $k_{\rightarrow}^{(i)}$ and $f_{i}^{*}$ respectively. }}
\label{f16}
\end{center}
\end{figure}

\begin{figure}[H]
\begin{minipage}{7cm}
\begin{center}
\includegraphics[height=6.5cm]{33st.eps}\\
(a)
\end{center}
\end{minipage}
\begin{minipage}{7cm}
\begin{center}
\includegraphics[height=6.5cm]{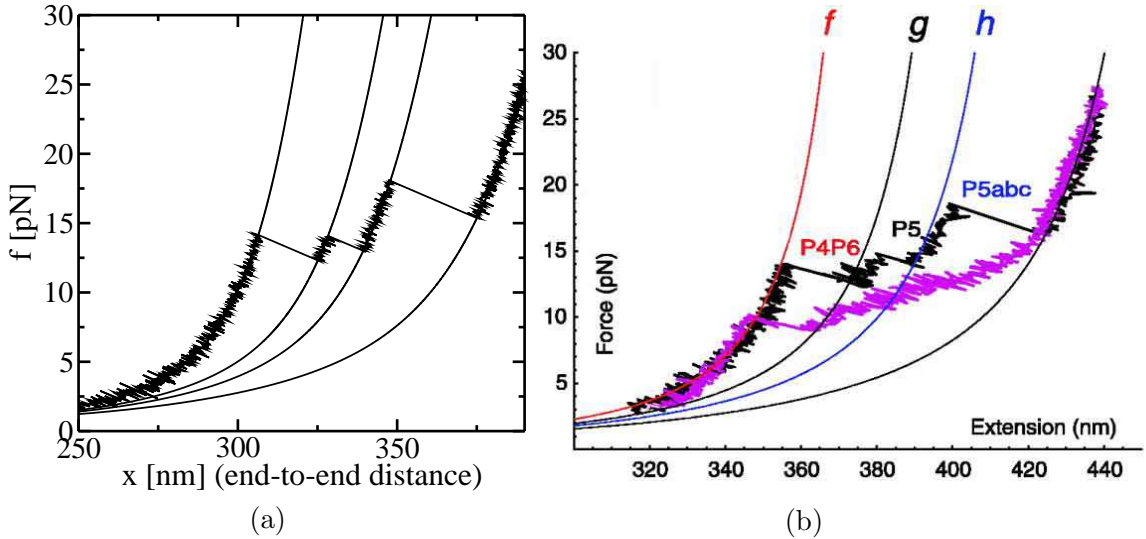}\\
(b)
\end{center}
\end{minipage}
\caption{\small{Comparison of FECs between model and experiments.
(a) Numerical simulations of the pulling process are at
$r=4\rm pN/s$ for three RNA domains. Simulations have been done with
the effective model \eq{s111} without refolding. Domains characterized
by $\tilde{x}_{1}^{(1)}=2.5\rm nm$ $\tilde{B}^{(1)}\ln(k_{o}^{(1)})=8.5k_{B}T$,
$\tilde{x}_{1}^{(2)}=2.5\rm nm$ $\tilde{B}^{(2)}\ln(k_{o}^{(2)})=8k_{B}T$,
$\tilde{x}_{1}^{(3)}=1.7\rm nm$ $\tilde{B}^{(3)}\ln(k_{o}^{(3)})=8.5k_{B}T$, where
the super-index refers to the index of the domain. The solid lines correspond to the WLC
force-extension curves. (b) Experimental FEC for the P4-P6
domain obtained in \cite{b11}. The solid lines correspond to WLC
curves for the handles linked to the RNA molecule. The lower curve
correspond to the refolding process that we do not consider
here. Figure taken from \cite{b11}.}}
\label{f17}
\end{figure}

\begin{figure}[H]
\begin{minipage}{7cm}
\begin{center}
\includegraphics[height=6.5cm]{hist.eps}\\
(a)
\end{center}
\end{minipage}
\begin{minipage}{7cm}
\begin{center}
\includegraphics[height=6.5cm]{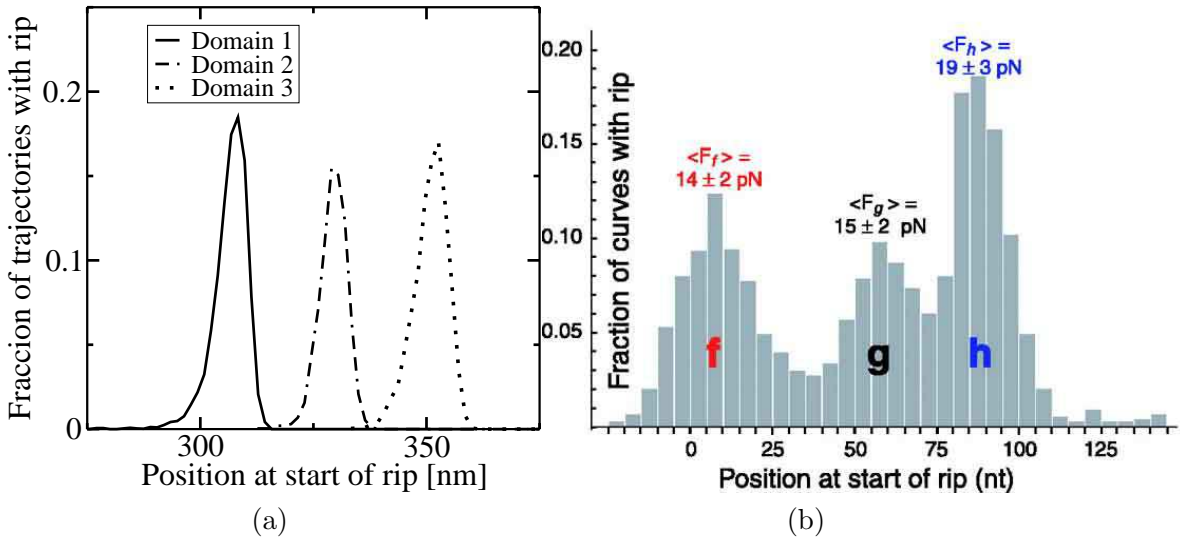}\\
(b)
\end{center}
\end{minipage}
\caption{\small{Comparison between model and experiments of the rip position 
distribution. (a) Histograms for the positions at the start of the
detected rips. They correspond to the three transitions observed in
Fig.~\ref{f17} (a). The parameters used in the simulation are
given in Fig.~\ref{f17} (a). (b) Experimental histograms of
rips detected in 732 unfolding curves of P4-P6 (Fig.~~\ref{f17} (b)). Figure taken from
\cite{b11}.}}
\label{f17b}
\end{figure}

For sake of clarity we will consider a sequential unfolding of a
multidomain RNA molecule. In general the unfolding of domains is a
hierarchical process not necessarily sequential. For instance, in
left panel of Fig.~\ref{f14}, once D1 has opened, either D2 and D3 can be
unfolded. However in our modelization we unfold sequentially the
domains D2 and D3. The motivation to consider this simplified model is
twofold. On the one hand, there are experimental results \cite{b11}
on the molecule L-21, a derivative of the Tetrahymena thermophila ribozyme, 
where the order of the opening of the different domains of the
molecule studied was never observed to change. On the other hand, a
main goal throughout this paper is to illustrate how the model for the
experimental setup previously introduced in
Secs.~\ref{setup}, \ref{twostates}, \ref{parts} can be generalized to
include complex RNA molecules (and not only hairpins) rather than
emphasizing details of the modeling of the RNA structure. With this
proviso, we then model the RNA molecule as an unidimensional chain of
single domains connected in series, each one represented as a
two-states model. For a $n$ domain system we have the F state, the UF
one, and $n$-1 intermediates, $I_{i}$, where $i$ is the index of the
intermediate (Fig.~\ref{f16}).
We simulate a pulling process without refolding using the effective
unfolding rate given in (\ref{s111}) for a molecule with three domains
in series. This system could represent the domain P4-P6 of the
molecule L-21, recently investigated \cite{b11}. In these experiments,
it is observed a sequential unfolding of the domains, even though
there are different unfolding pathways because not all the
intermediates are seen in each trajectory (sometimes two consecutive 
domains open simultaneously). The most frequently
observed pathway contains three transitions corresponding to the
consecutive opening of the domains P4P6, P5 and P5abc. In
Fig.~\ref{f17} we show the FEC of a 3-domain RNA system and in Fig.~\ref{f17b}
the histograms for the starting position of the rips detected. The results shown  in Figs.~\ref{f17} (a) \ref{f17b} (a) have
been obtained by doing a numerical simulation of a pulling experiment
using the parameters for the handles and the bead given in
Table~\ref{table1}. The kinetic parameters of each RNA domain are
given in Fig.~\ref{f17}. In panels (b) of these figures are shown the 
 experimental results \cite{b11}.

For the third domain, that corresponds to the well known domain P5abc,
we use the values of the parameters $\tilde{x}_{1}^{(3)}$ and $\tilde{B}^{(3)}\ln(k_{o}^{(3)})$ obtained in \cite{b10}. We choose the parameters for the
other domains in order to qualitatively reproduce the experimental
results for the unfolding trajectories in \cite{b11} shown in the
right panel of Fig.~\ref{f17}. The histograms for the positions at the
start of the rips detected obtained from these values of parameters
are different from the experimental ones (Fig.~\ref{f17b}). The main
difference comes from the amplitude of the fluctuations of the
position where each domain opens, that is smaller in simulations
results as compared to experimental results~\footnote{Note that in 
experimental results the distances are given in units of nucleotides
but our results are in units of nanometers. In the totally extended
form of the polymer the conversion unit is $0.59\rm nm$ per
nucleotide.}. Several reasons can explain this disagreement. First,
there are strong drift effects in the machine that introduce
instrumental noise. Second, no two pulled molecules are ever identical
(disparity of the attachments, existence of more than one tether on
the beads that can influence force measurements). A dispersion in the
population of molecules is always another source of noise.  Third, the
RNA molecule is not just composed by a series of domains, but there
are other regions (some bases) that do not belong to any domain. These
regions can contribute differently to increase the length of the rips,
a source of randomness for the position of the start of the rips. Last
but not least, we cannot exclude that the kinetic model we are
considering is too simple to explain the unfolding of these
domains. It is known that complex RNA structures show characteristic
FECs that cannot be usually interpreted in terms of the successive opening of
native domains, because of the existence of long-lived intermediates
including non-native helices \cite{b32}.

The most important difference between the analysis of a single
barrier (see Appendix~\ref{app_singledomain}) and the present study of a 
succession of $n$ domains  is that the force does not reach the domain
 $i$ until the previous domain $i$-1 has opened. 
Then the domain $i$ starts to be pulled only at a force larger than 
$f_{i}^{s}$, which in our approximation is given by: 
\be
f_{i}^{s}=f_{i-1}^{*}-k_{b}\tilde{x}_{m}^{(i-1)}~.
\label{m13}
\ee 
where the parameters and functions with index $i$ refer to the
domain $i$.  Let us define the quantity $C_{i}$ as:
\be
C_{i}=\frac{k_{B}T}{r\tilde{x}_{1}^{(i)}}k_{\rightarrow}^{(i)}(f_{i}^{s})~.
\ee
 The average value over different trajectories of this quantity
$\langle C_{i}\rangle$ is a measure of the probability of opening the
domain $i$ just after the domain $i$-1 has been opened. According to the
value of $\langle C_{i}\rangle$ we can distinguish three different
regimes:
\begin{enumerate}
\item
$\langle C_{i}\rangle<<1$.  Most of trajectories show two separated
transitions (rips) for the opening of the domain $i$-1 and $i$,
because at the typical value of $f_{i}^{s}$ there is a low probability
of opening the domain $i$. It is then possible to treat the domain $i$
independently of the $i$-1, as a single domain, using (\ref{m7}) to
analyze the distribution of breakage forces.
\item
$\langle C_{i}\rangle>>1$. The probability of opening the domain $i$ at $f_{i}^{s}$ is large. Therefore most of the time one observes a single transition (rip) for the opening of both domains and the intermediate state $I_{i-1}$ is hardly observable. In this case it is not possible to obtain information about the domain $i$. 
\item
$\langle C_{i}\rangle\approx1$. This is the intermediate case between
the two previous ones. We expect to observe trajectories with a single
transition (1 rip) for the opening of the domains $i$ and $i$-1 and
other ones with two separated transitions (2 rips). In this case, to
obtain kinetic information of domain $i$ from the analysis of the
distribution of breakage force, $f_{i}^{*}$, we need to recalculate the
distribution of the breakage force as shown in appendix~\ref{PF3} or to
work with the distribution of $f_{i}^{*}$ conditioned to the fact the
domain $i-1$ has been opened at a force smaller than a given value.
\end{enumerate}

We focus on the study of the regime 3 considering two different
two-domain molecules coupled to the system described in
Sec.~\ref{setup} with parameters given in Table~\ref{table1}. The
system is pulled at $r=4\rm pN/s$. The first domain is the same for
both molecules and its kinetics parameters are
$\tilde{x}_{1}^{(1)}=2.5\rm nm$ and $\tilde{B}^{(1)}\ln(k_{o}^{(1)})=8.5k_{B}T$; the
second domain is different for the two molecules, but both have
$\langle C_{2}\rangle$ of order of 1, so they are in regime 3. In
order to get the kinetics parameters for the second domain, we will
use two different techniques:
\begin{itemize}
\item
In appendix~\ref{PF3} we compute the distribution of breakage forces
for a domain $i$ in regime 3 (\ref{m17}), as a function of the
kinetics parameters of domain $i$ and the previous one $i-1$. This
technique uses the expression (\ref{m17}) to extract kinetic
information for the second domain from the kinetics parameters of the
first domain. The method consist in first building an histogram of the
breakage forces for the second domain, using the results from all the
trajectories\footnote{\label{foot14}For the trajectories where only a
single transition is observed for the opening of the first and second
domains, it is possible to compute the breakage force for the second
domain $f^{*}_2$ as:
$f^{*}_2=f^{*}_{1}-k_{b}\tilde{x}_{m}^{(1)}$.}. Then we fit the
histogram to the distribution (\ref{m17})\footnote{\label{foot15}We
truncate the series at certain $k$ once we find convergence.} to get
the kinetics parameters for the second domain. The results obtained
are shown in Fig.~\ref{f19}.

\begin{figure}[H]
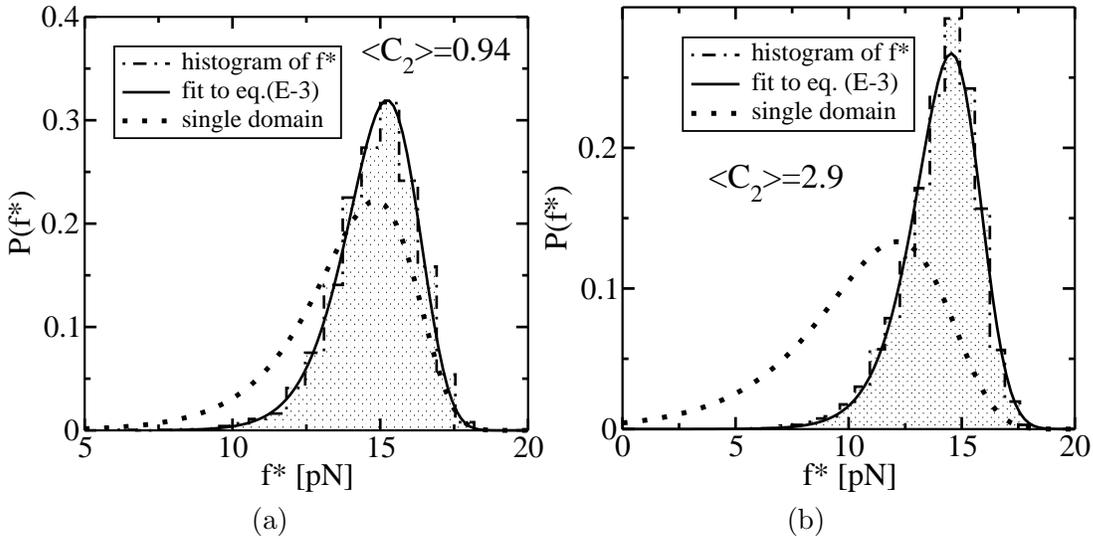

\begin{minipage}{7cm}
\begin{center}
%FEC1r.eps
\includegraphics[height=6.5cm]{b1pr.eps}\\
(a)
\end{center}
\end{minipage}
\begin{minipage}{7cm}
\begin{center}
%FEC1p.eps
\includegraphics[height=6.5cm]{b2pr.eps}\\
(b)
\end{center}
\end{minipage}
\caption{\small{Histogram of breakage forces for the second domain of
a two domain system characterized by given values of
$\tilde{x}_{1}^{(2)},\tilde{B}^{(2)}\ln(k_{o}^{(2)})$. The continuous line is the best fit to
(\ref{m17}), truncating the series at a value $k=k^*$ where convergence is
achieved. The dotted line shows the distribution of breakages forces for a
single domain for the real values of $\tilde{x}_{1}^{(2)}$ and  $\tilde{B}^{(2)}\ln(k_{o}^{(2)})$. (a): System with $\tilde{x}_{1}^{(2)}=2.5\rm nm$ and
$\tilde{B}^{(2)}\ln(k_{o}^{(2)})=8k_{B}T$. Series summation was truncated at $k^*=2$.
The fit gives $\tilde{x}_{1}^{(2)}=2.6\pm0.1\rm nm$ and
$\tilde{B}^{(2)}\ln(k_{o}^{(2)})=8.2\pm0.3k_{B}T$ in agreement with the correct
values. The average value for the parameter $C_{i}$ for this domain is
$<C_{2}>=0.94$. These are results obtained from 1000 pulls.
(b):  System with $\tilde{x}_{1}^{(2)}=1.5\rm nm$ and
$\tilde{B}^{(2)}\ln(k_{o}^{(2)})=4k_{B}T$. Series summation was truncated at $k^*=3$.
The fit gives $\tilde{x}_{1}^{(2)}=1.6\pm0.1\rm nm$ and
$\tilde{B}^{(2)}\ln(k_{o}^{(2)})=4.3\pm0.3k_{B}T$ in agreement with the correct
values. The average value for the parameter $C_{i}$ for this domain is
$<C_{2}>=2.9$. These are results obtained from 1000 pulls.}}
\label{f19}
\end{figure}

\item
The second technique consist on working with the probability
distribution that the domain $i$ opens at a force $f_{i}^{*}$
conditioned to the fact that the previous domain opened at a force
$f_{i-1}^{*}$ smaller than a given force $f_{l}$,
$\rho(f_{i}^{*}|f_{i-1}^{*}<f_{l})$. Considering small values of
$f_{l}$, the distribution $\rho(f_{i}^{*}|f_{i-1}^{*}<f_{l})$ gets
closer to the distribution of a single domain (\ref{m7}). For instance
if we consider $f_{l}<\tilde{f}$, where $\tilde{f}$ is the minimal force
at which there is no probability of unfolding the domain $i$ at the
given $r$~\footnote{ The force $\tilde{f}$ represents the lower limit force
value below which the distribution of breakage force goes to zero.} 
, the conditioned
distribution overlaps with the distribution for a single domain. To
compute $\rho(f_{i}^{*}|f_{i-1}^{*}<f_{l})$, we do histograms of
breakage forces for the set of trajectories that verify
$f_{i-1}^{*}<f_{l}$. Starting with a certain value of $f_l$ we build the
histogram and do the fit to (\ref{m7}) to get the kinetics parameters,
$\tilde{x}_{1}^{(i)}$ and $\tilde{B}^{(i)}\ln(k_{o}^{(i)})$. Then we
repeat the process decreasing the value of $f_l$ until the parameters
obtained from the fit converge to a given value; in this regime of
values of $f_l$ the domain $i$ is not influenced by the previous domain,
and one gets the right values for the kinetics parameters. The drawback
of this technique is that for $f_{l}$ too small the number of useful
trajectories quickly decreases, and one needs many more pulls to be able
to build an histogram. In the following figure \ref{f22} we show our
results for $\rho(f_{2}^{*}|f_{1}^{*}<f_{l})$, for the two molecules
considered before.

\begin{figure}[H]
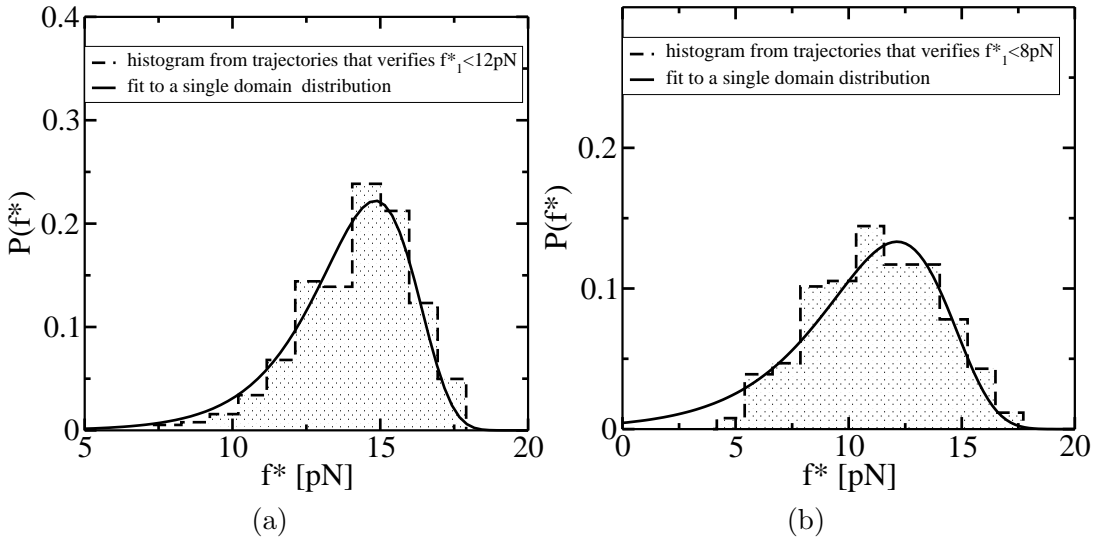

\begin{minipage}{7cm}
\begin{center}
%FEC1r.eps
\includegraphics[height=6.5cm]{b1.2p.eps}\\
(a)
\end{center}
\end{minipage}
\begin{minipage}{7cm}
\begin{center}
%FEC1p.eps
\includegraphics[height=6.5cm]{b2.2p.eps}\\
(b)
\end{center}
\end{minipage}
\caption{\small{Histogram of breakage forces for the second barrier of
a two domain RNA molecule for the set of trajectories that verify
$f_{1}^{*}<f_l$ (the first domain opens at a force smaller than
$f_l$). The continuous line corresponds to the distribution of
breakages forces for a single domain for the real values of
$\tilde{x}_{1}^{(2)}$ and $\tilde{B}^{(2)}\ln(k_{o}^{(2)})$. (a): Same
parameters as in Fig.~\ref{f19} (a). System
characterized by $\tilde{x}_{1}^{(2)}=2.5\rm nm$ $\tilde{B}^{(2)}\ln(k_{o}^{(2)})=8k_{B}T$,$f_l=12\rm pN$. The fit to (\ref{m7}) gives
$\tilde{x}_{1}^{(2)}=2.5\pm0.1\rm nm$ and $\tilde{B}^{(2)}\ln(k_{o}^{(2)})=8.0\pm0.4k_{B}T$.  Histograms were obtained from 3000 pulls
and 392 pulls verify $f_{1}^{*}<12\rm pN$. (b): Same
parameters as in Fig.~\ref{f19} (b). System
characterized by $\tilde{x}_{1}^{(2)}=1.5\rm nm$, $\tilde{B}^{(2)}\ln(k_{o}^{(2)})=4k_{B}T$, $f_l=8\rm pN$. The fit to (\ref{m7}) gives
$\tilde{x}_{1}^{(2)}=1.5\pm0.1\rm nm$ and $\tilde{B}^{(2)}\ln(k_{o}^{(2)})=3.7\pm0.3k_{B}T$. Histograms were obtained from 15000 pulls
and 204 pulls verify $f_{1}^{*}<8\rm pN$.}}
\label{f22}
%\end{center}
\end{figure}
\end{itemize}

We conclude that in order to obtain kinetic information for a domain in
regime 3 both techniques are complementary. The first technique has 
the disadvantage that it requires the knowledge of the kinetic parameters 
of the previous domain. The method to extract information 
about the second domain is to start with the
analysis of the first domain (that is not blocked by any domain) and
going forward following the sequential order in which domains open. On the
other hand, the problem of the second technique is that when considering
small values of $f_{l}$ the number of useful trajectories quickly decreases, 
and one needs a large number of pulls to be able to build an
histogram. Depending on the experimental conditions one can
decide which technique is the best to apply.

\section{Conclusions} 
\label{conclusions}
The recent fast development of nanotechnologies allow scientists to
investigate the physical behavior of complex biomolecules. Of particular
importance are those physical processes in the nanoscale where the typical values of the
energies involved are several times $k_BT$.  In such regime fluctuations and
large deviations from the average behavior are important and deserve a
careful investigation as they can contribute a lot to the understanding
of thermal processes in small systems.
RNA pulling experiments offer an excellent framework to address such
questions as RNA molecules can be small enough for stochastic fluctuations
be observable and measurable. 

An extremely useful technique to manipulate individual molecules are
optical tweezers which cover a range of forces 1-100pN that is relevant
for many biological processes. A full understanding about
how to extract accurate physical information from such experiments is
therefore of great importance. 
The present work represents an attempt in that direction. At present it
is not yet possible to unfold individual RNA molecules without attaching
some polymer handles at their extremes, therefore all RNA pulling experiments are
carried out with a system larger than the individual ``naked'' RNA
molecule. This system includes the RNA molecule, the polymer handles
and the bead in the optical trap. In order to extract accurate physical
information regarding the RNA molecule, a global treatment of the
whole system is necessary.

In this paper we analyzed the minimal system required to interpret the
data extracted from RNA pulling experiments. We did not include any
details regarding the response of the machine or a realistic and
accurate modelization of the structure of the RNA molecule. On the
contrary, we have focused on those thermodynamic and kinetic aspects of
pulling experiments by considering the transmission of the force on the
RNA molecule induced by the the bead and handles. A key part of our
treatment is a proper consideration of the ensemble that is relevant in pulling 
experiments. While the end-to-end distance (between the bead and the
micropipette) and the force are variables that fluctuate,
the total end-to end distance $X_T$ (Fig.~\ref{f1}) does not. The thermodynamic
potential in such ensemble is the key quantity that allows us to extract
accurate knowledge of the influence of these external parts (beads and
handles) on the thermodynamic and kinetic behavior of the RNA molecule.

In Sec.~\ref{twostates} we introduce the appropriate thermodynamic
potential by focusing the analysis on single domain RNA molecules that
show a highly cooperative folding-unfolding behavior.  In
Sec.~\ref{thermo} we analyzed the thermodynamics of the whole system by
doing a partition function analysis that includes all parts of the setup
previously described in Sec.~\ref{parts}. Four are the most important
results in Sec.~\ref{thermo}: a) we get an explicit expression \eq{w1}
for the transition force $F^c$ as well as we are able to reconstruct the
thermodynamic force-extension curve (TFEC) from the knowledge of the
parameters of the model, see \eq{new1}; b) The different contributions
to the total reversible work \eq{w4a}, coming out from the different
parts of the system (bead, handles and RNA molecule), have been analyzed
\eqqq{w4b}{w4c}{w4d}. A comprehensive summary of these results is shown
in Fig.~\ref{f4}; c) A relation between the unfolding free-energy of
the molecule $\Delta G_0$ and the area under the force rip $W_{\rm
rip}^c$ has been given in \eq{r4}. d) Finally the dependences of the
free-energy contributions to the total reversible work across the
transition were analyzed as a function of the stiffness of the trap and
the ratio between the contour and persistence lengths of the polymer
handles (Fig.~\ref{f10b} and Table~\ref{tablew}).
 Taken together all these results
establish a framework to infer thermodynamic properties of the RNA
molecule from the experimental data. Moreover, they also allow us to
understand under which conditions (parameters for the bead and handles)
it is more reliable to get estimates for these properties.

From the thermodynamics to the kinetics we verify in
Sec.~\ref{simulation} that the model studied qualitatively reproduces
the results reported from experiments (Figs.~\ref{f7} and \ref{f8b}) 
doing a numerical simulation of a pulling experiment. In
Sec.~\ref{fraction} we obtain some interesting results for
other quantities that are amenable to experimental checks. In particular, we
find a generic relation between the fraction of molecules that unfold
(refold) at least twice during the unfolding (refolding) and the mean
dissipated work. Interestingly this relation is valid beyond the linear
response regime where the dissipated work does not increase linearly
with the pulling speed.  This relation could allow us to extract
the reversible work for the unfolding process by using data extracted
from non-equilibrium pulling experiments.  This procedure is reminiscent 
of other techniques,
recently applied to RNA pulling experiments~\cite{Jarzynski}, based on the
Jarzynski equality or similar relations (for a recent review,
see~\cite{Ritort}). Moreover we have shown a symmetry property that relates
these fractions for the forward and reverse processes. How general this
result is in general
transition state theory \cite{Chandler} (i.e. beyond the case of a cooperative two-states
system) remains an interesting open question.

In order to stress the adaptability and feasibility of our model to
describe more complex type of molecules we consider in
Sec.~\ref{domains} the unfolding of a large RNA molecule made out of
different domains that unfold sequentially. The unfolding of these
domains is controlled by $Mg^{2+}$ tertiary interactions which induce
large energy barriers, thereby a refolding event (while the molecule is
pulled) is not observed at experimental conditions.  Although our study
is not complete for such type of molecules (the assumption of a
sequential unfolding may not consider other possible unfolding
pathways) it is instructive to see that by  modifying only the model for the RNA
molecule we are still capable of qualitatively reproducing several experimental
results as shown in Figs.~\ref{f17} and \ref{f17b}. Finally we discuss
possible ways to extract information about the kinetics of a single
domain from the analysis of the breakage force distribution in a regime
where the distribution of the breakage force for a domain depends
on the presence of a previous domain. 

Many aspects of RNA pulling experiments are still open, among them would
be interesting to extend these considerations to include more complex
effects induced by the response of the machine, test experimentally some
of the results predicted in this work for the fraction of unfolded
events and also a detailed investigation of the kinetics of the folding
process (rather than the unfolding) in the presence of force, a process
for which we still lack an understanding. Several of these aspects
will be addressed in the near future.

\section*{Acknowledgments}
We thank C. Bustamante, J. Liphardt, I. Tinoco and S. Smith for insightful
discussions.  We also thank I. Pagonabarraga and G. Franzese  for 
discussions as well as a
critical reading of the manuscript. M. Ma\~nosas has been supported by U.B
Grant and F. Ritort has been supported by the David and
Lucile Packard Foundation, the European community (STIPCO network), the
Spanish research council (Grant BFM2001-3525) and the Catalan
government.

\appendix

\renewcommand{\theequation}{A-\arabic{equation}}
\setcounter{equation}{0}
\section{Partition function in mixed ensemble \label{append_part}} 
The partition function, $Z(X_{T})$, for the system described in Fig.~\ref{f1},
gives the free energy $G_{X_{T}}$ as well as other relevant thermodynamic
properties. The state of the system is defined by the externally
controlled variables $X_T$, $T$ and $P$. The last two, $T$ and $P$, are
always kept at a constant value so we can ignore them throughout the
paper. The partition function for this one-dimensional system can be
written as the convolution of the contributions coming out from the different elements~\footnote{\label{foot16a}We restrict the configurational space to positive values of the variables $x_{\alpha}$, with $\alpha=h_{1},h_{2},r,b$. The reason of taking this simplification is because  when considering positive values of the control parameter $X_{T}$ the configurations with negative values of some $x_{\alpha}$ have practically no weight.}:
\begin{eqnarray}
Z(X_{T})&=&C\int_{0}^{L_{1}}dx_{h_{1}}\int_{0}^{L_{2}}dx_{h_{2}}\int_{0}^{\infty}dx_{b}\int_{0}^{L_{r}}dx_{r}\Big{[}Z^{h_{1}}(x_{h_{1}})Z^{h_{2}}(x_{h_{2}})Z^{s}(x_{b})Z^{r}(x_{r})\nonumber\\
& & \times {\delta}(X_{T}-(x_{h_{1}}+x_{h_{2}}+x_{b}+x_{r}))\Big{]}~,
\label{Z1}
\end{eqnarray}
where $Z^{\alpha}(x_{\alpha})$ is the partition function distribution of
the element $\alpha$, with ${\alpha}=h_{1},h_{2},r,b$. The lengths
$L_{1}$, $L_{2}$ and $L_{r}$ are the contour lengths of the handles 1, 2
and the single stranded RNA (ssRNA) respectively. The constant $C$ is a
normalization factor.  The distribution $Z^{\alpha}(x_{\alpha})$ for the element $\alpha$ is computed as: 
\be
Z^{\alpha}(x_{\alpha})=g_{\alpha}(x_{\alpha})e^{-{\beta}E_{\alpha}(x_{\alpha})}=e^{-{\beta}G_{\alpha}(x_{\alpha})}~,
\label{Z2}
\ee 
with $\beta=\frac{1}{k_{B}T}$. The functions $E_{\alpha}(x_{\alpha})$ and 
$G_{\alpha}(x_{\alpha})$ are the energy (or
enthalpy) and the Gibbs free energy of the element $\alpha$
respectively. Both are related by
$G_{\alpha}(x_{\alpha})=E_{\alpha}(x_{\alpha})-TS_{\alpha}(x_{\alpha})$
where $S_{\alpha}(x_{\alpha})$ is the entropy,
$S_{\alpha}(x_{\alpha})=k_{B}\ln (g_{\alpha}(x_{\alpha}))$ and
$g_{\alpha}(x_{\alpha})$ is the density of states. We now
compute the free energy $G_{\alpha}$, of each of the different elements
at fixed value of $x_{\alpha}$:
\begin{itemize}
\item
Bead trapped in a potential well: As the $V_{b}(x)$ is the 
potential of mean-force for the bead in the trap along the reaction
coordinate (see footnote \ref{foot2}) we can write, 
\be
Z^{s}(x_{b})=e^{-{\beta}V_{b}(x_{b})}~.
\label{Z3}
\ee
\item Handles:
We use the fact that the difference of free energy between the state defined with $x=0$ and the one with $x=x_{h_{i}}$ is equal to the reversible work performed by stretching the handle from $x=0$ to $x=x_{h_{i}}$, 
\be
G_{h_{i}}(x_{h_{i}})=\int_{0}^{x_{h_{i}}}dxf_{h_{i}}(x)=W_{h_{i}}(x_{h_{i}})~,\;\;{\rm
for}\;\;i=1,2
\label{Z4}
\ee
where $f_{h_{i}}(x)$ is the thermodynamic force-extension curve (TFEC) of the handle $i$~\footnote{\label{foot16} Which variables are controlled is a relevant choice in single molecule pulling experiments. In contrast with macroscopic systems, where all the ensembles are equivalent \cite{b13bibi}, FECs depend on the particular ensemble considered. Eq.~\eq{Z4} has been defined for the isometric ensemble. The isometric TFEC is the thermodynamic curve corresponding to a system in the ensemble where the end-to-end distance $x$ is held fixed, and is given by the mean force as a function of $x$, $\langle f \rangle (x)$. While the isotensial TFEC is the TFEC resulting of working in the force ensemble, $\langle x \rangle (f)$. In general both TFEC differ \cite{b13bibi}, but in this analysis  we consider that the handles and the RNA molecule are long and flexible enough to have an identical isometric and isotensional TFEC that we call $f_{\alpha}(x_{\alpha})$ with $\alpha=h_{1},h_{2},r$. This allow us to use the extrapolation expression (\ref{p2}) (or the one given in \cite{b25}) for the function $f_{\alpha}(x_{\alpha})$ when using the WLC model to describe the polymer behavior.}. We get \be
Z^{h_{i}}(x_{h_{i}})=e^{-{\beta}W_{h_{i}}(x_{h_{i}})}~. 
\label{Z5}
\ee
\item
RNA: The partition function $Z^{r}$ can be divided in two parts, one
corresponding to the F state ($\sigma=0$) and the other to the UF state
($\sigma=1$).  In the present analysis we are considering that the F state is
represented by a single configuration while the UF states are represented
by a continuous set of configurations corresponding to the difference
extensions of the ssRNA (Fig.~\ref{f3} (b)). Therefore $Z^{r}$ is made up of
two terms: a singular contribution that comes from the F state
($\sigma=0$) represented by a delta function and a continuous
contribution that comes from the UF state ($\sigma=1$). We take the F
state as the reference state with zero free energy. The free energy of
the UF state has two terms: the free energy at zero force,
${\Delta}G^{0}$, plus the corresponding loss of entropy due to the
stretching:
\be
Z^{r}(x_{r})=Z(x_{r},\sigma=0)+Z(x_{r},\sigma=1)=\delta(x_{r})+C_{r}e^{-{\beta}({\Delta}G^{0}+W_{r}(x_{r}))}~~,
\label{Z6b}
\ee
where $W_{r}(x_{r})$ is computed as in (\ref{Z4})
\be
W_{r}(x_{r})=\int_{0}^{x_{r}}dxf_{r}(x)~~,
\label{Z6}
\ee
being $f_{r}(x)$  the TFEC of the ssRNA
polymer. The probability $P(\sigma)$ for the RNA molecule to be in the state 
$\sigma$ is given by 
$P(\sigma)\propto\int_{0}^{L_{r}}dx_{r}Z(x_{r},\sigma)$. To compute $C_{r}$ we use that the RNA molecule at zero force satisfies   
\be
{\Delta}G^{0}=-k_BT\ln (\frac{P(\sigma=1)}{P(\sigma=0)})
\label{new}
\ee
and substituting \eq{Z6b} we obtain,
\be
C_{r}=\frac{1}{\int_{0}^{L_{r}}dxe^{-{\beta}W_{r}(x)}}~.
\label{Z7}
\ee
\end{itemize}

\begin{figure}[H]
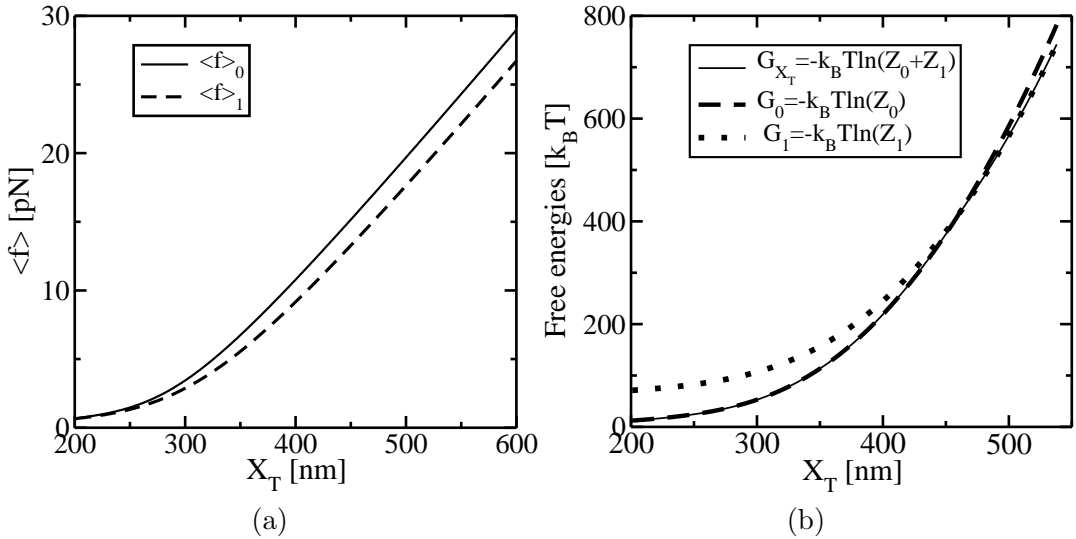

\begin{minipage}{7cm}
\begin{center}
\includegraphics[height=6.5cm]{f0f1.eps}\\
(a)
\end{center}
\end{minipage}
\begin{minipage}{7cm}
\begin{center}
\includegraphics[height=6.5cm]{G01.eps}\\
(b)
\end{center}
\end{minipage}
\caption{\small{We consider a system with the parameters given in Tables~\ref{table1} and ~\ref{table1b}. From the partition function analysis we compute: (a) The two branches $\langle f\rangle_{\sigma}$, corresponding to the thermodynamic forces acting upon the system for a given $\sigma$ RNA state as a function of $X_T$. (b) The free energy $G_{X_{T}}$ and the free energy of each branch $\sigma$, $G_{\sigma}$, as a function of $X_{T}$.}}
\label{lambda}
\end{figure}

Adding the different contributions we get:
\begin{eqnarray}
Z(X_{T})&=&C\int_{0}^{L_{1}}dx_{h_{1}}\int_{0}^{L_{2}}dx_{h_{2}}\int_{0}^{\infty}dx_{b}\int_{0}^{L_{r}}dx_{r}\Big{[}e^{-{\beta}(W_{h_{1}}+W_{h_{2}}+V_{b})}\nonumber\\
&\times &[\delta(x_{r})+C_{r}e^{-{\beta}({\Delta}G^{0}+W_{r})}]{\delta}(X_{T}-(x_{h_{1}}+x_{h_{2}}+x_{b}+x_{r}))\Big{]}~.
\label{Z8}
\end{eqnarray}
We now separate \eq{Z8} in two contributions coming from the F and the
UF states. By using the integral representation of the delta function, 
\begin{eqnarray}
\delta(x)=\frac{1}{2\pi}\int_{-\infty}^{\infty}\exp(i\lambda x)d\lambda
\label{Z8b}
\end{eqnarray}
we get
\begin{eqnarray}
Z(X_{T})=Z_0(X_{T})+Z_1(X_{T}) \;\; {\rm with}
\label{Z9a}     
\end{eqnarray}
\begin{eqnarray}
Z_0(X_{T})=\frac{C}{2\pi}\int_{-\infty}^{\infty}d{\lambda}e^{(i{\lambda}X_{T}+g_{0}({\lambda}))}\;\;{\rm
and} \;\;\; Z_1(X_{T})=\frac{C}{2\pi}\int_{-\infty}^{\infty}d{\lambda}e^{(i{\lambda}X_{T}+g_{1}({\lambda}))},
\label{Z9}     
\end{eqnarray}
where the functions $g_0$ and $g_1$ are given by
\begin{eqnarray}
g_{0} = \log\Big{[}\int_{0}^{L_{1}}dx_{h_{1}}\int_{0}^{L_{2}}dx_{h_{2}}\int_{0}^{\infty}dx_{b}[e^{-{\beta}(W_{h_{1}}+W_{h_{2}}+V_{b})}e^{-i{\lambda}(x_{h_{1}}+x_{h_{2}}+x_{b})}]\Big{]}~,
\label{Z10}
\end{eqnarray}
\begin{eqnarray}
g_{1}& =&\log\Big{[}\int_{0}^{L_{1}}dx_{h_{1}}{\int}_{0}^{L_{2}}dx_{h_{2}}{\int}_{0}^{\infty}dx_{b}{\int}_{0}^{L_{r}}dx_{r}[C_{r}e^{-{\beta}(W_{h_{1}}+W_{h_{2}}+V_{b}+{\Delta}G^{0}+W_{r})}\nonumber\\
& &e^{-i{\lambda}(x_{h_{1}}+x_{h_{2}}+x_{b}+x_{r})}]\Big{]}~.
\label{Z11}
\end{eqnarray}
Eqs.~(\ref{Z9}) for $Z_0$ and $Z_1$ are integrals respect to $\lambda$ of
an exponential with an argument that is extensive with the size of the
system~\footnote{\label{foot17}By size we mean the length of the handles as well as the
length or molecular weight of the RNA molecule. In general to apply the saddle point approximation we require
that the energies of the different elements of the system (bead, handles
and molecule) are several times $k_{B}T$.}. Therefore if the system is big enough,
the saddle point approximation is valid and becomes exact in the
thermodynamic limit. As a check we have verified that the results from
the saddle point approximation and the exact numerical integration
of the partition function are in pretty good agreement for the system with parameters given in Tables~\ref{table1} and ~\ref{table1b}.  Applying the
saddle point technique, one is led to extremize the arguments of the
exponentials  with respect to all the variables of integration. In this way we obtain:
\begin{eqnarray}
\frac{dg_{\sigma}}{dx_{\alpha}}\Big{|}_{x_{\alpha}=\tilde{x}_{\alpha}^{\sigma}}=\tilde{\lambda}_{\sigma} ~~{\rm{ with}}~~ \sigma=0,1 ~~{\rm{ and}} ~~\alpha=h_{1},h_{2},r,b~,
\label{Z12}
\end{eqnarray}
where $\tilde{x}_{\alpha}^{\sigma}$ corresponds to the value of the  variable $x_{\alpha}$  when the RNA molecule is in the state $\sigma$ that extremizes the argument of the exponential. We have two branches corresponding to the situations where the RNA is folded ($\sigma=0$) and where the RNA is unfolded ($\sigma=1$). We use the super-index $\sigma$ to denote each branch. Eq. (\ref{Z12}) tells that the integration variable $\lambda$ plays the role of the thermodynamic force, so the $\tilde{\lambda}_{\sigma}$ corresponds to the mean force acting upon the system for the branch $\sigma$
and for a fixed value of $X_T$ called $\langle f\rangle_{\sigma}$. 
Eq.\eq{Z12} can be written as 
\begin{eqnarray}
f_{b}^{0}(\tilde{x}_{b}^{0})&=&f_{h_{1}}^{0}(\tilde{x}_{h_{1}}^{0})=f_{h_{2}}^{0}(\tilde{x}_{h_{2}}^{0})=\langle f\rangle_{0}~,\nonumber\\
f_{b}^{1}(\tilde{x}_{b}^{1})&=&f_{h_{1}}^{1}(\tilde{x}_{h_{1}}^{1})=f_{h_{2}}^{1}(\tilde{x}_{h_{2}}^{1})=f_{r}^{1}(\tilde{x}_{r}^{1})=\langle f\rangle_{1}~,
\label{Z13}
\end{eqnarray}
where the force  $f_{\alpha}^{\sigma}=\langle \frac{dW_{\alpha}^{\sigma}(x)}{dx}\rangle$ is the mean force acting upon the element $\alpha$ at fixed $x_{\alpha}=\tilde{x}_{\alpha}^{\sigma}$ for the branch $\sigma$. In Fig.~\ref{lambda} (a) we show the two branches $\langle f\rangle_{\sigma}$ as a function of $X_T$ for a system with parameters given in Tables~\ref{table1} and ~\ref{table1b}. The transition from the F-UF states is the jump from one branch to the other.

Hence we have obtained that the values of the arguments for which the contribution to the partition function is maximum corresponds to the equilibrium values at a given $X_T$:
\be
Z(X_{T})=Z_0(X_{T})+Z_1(X_{T})~, \label{ar1a} 
\ee
\be
Z_{0}(X_{T})\approx \exp\Bigl[-{\beta}(W_{h_{1}}(\langle x_{h_{1}}\rangle_{0})+W_{h_{2}}(\langle x_{h_{2}}\rangle_{0})+V_{b}(\langle x_{b}\rangle_{0}))\Bigr]~,\label{ar1b}
\ee
\be
Z_{1}(X_{T})\approx \exp\Bigl[-{\beta}(W_{h_{1}}(\langle x_{h_{1}}\rangle_{1})+W_{h_{2}}(\langle x_{h_{2}}\rangle_{1})+V_{b}(\langle x_{b}\rangle_{1})+{\Delta}G^{0}+W_{r}(\langle x_{r}\rangle _{1}))\Bigr]~.
\label{ar1c}
\ee
where we have neglect the subdominant contributions. The $\langle x_{\alpha}\rangle_{\sigma}$ correspond to the mean value of $x_{\alpha}$ for the branch $\sigma$ and for a fixed value of $X_{T}$.  In Fig.~\ref{lambda} (b) we show the results for the free energy of the system with parameters given in Tables~\ref{table1} and ~\ref{table1b} as a function of $X_T$, 
\be
G_{X_{T}}=-k_{B}T\ln (Z(X_{T}))~~,
\label{Gxt}
\ee
and also the free energies of the system for each branch $\sigma$,
\be
G_{\sigma}=-k_{B}T\ln (Z_{\sigma}(X_{T}))~~.
\label{Gsig}
\ee
The free energy of the system $G_{X_{T}}$ changes from one branch to the other at $X_{T}^{c}$, when both states are equal probable, i.e  $G_{0}=G_{1}$.

\renewcommand{\theequation}{B-\arabic{equation}}
\setcounter{equation}{0}
\section{Computation of the folding and unfolding rates in the mixed ensemble\label{append_foldunfold}} 
We model the kinetics of the folding-unfolding of RNA as a Kramers
activated process characterized by the following transitions rates:
\begin{eqnarray}
k_{\rightarrow}(X_{T})=k_{0}\exp[-\beta B(X_{T})]\nonumber\\
k_{\leftarrow}(X_{T})=k_{0}\exp[\beta(-B(X_{T})+\Delta G(X_{T}))]~,
\label{ssb1}
\end{eqnarray}
where $k_{0}$ is an attempt frequency that depends on the shape of the
free energy landscape, on the molecular damping and on the natural
frequency of the hydrogen bond oscillations \cite{b12}. The functions
$\Delta G(X_{T})$ and $B(X_{T})$ are the difference of free energy
between the F and UF states and the height of the kinetic barrier
located between them (Fig.~\ref{f3} (a))~\footnote{Note that the physical meaning of $\Delta G(X_{T})$ is completely different from $\Delta G_{X_{T}}$ (see (\ref{w6})). The latter corresponds to the free energy difference of the global system between two different values of $X_{T}$.}. Using the results obtained
from the partition function analysis we can write $\Delta G(X_{T})$ as:
 
\begin{eqnarray}
{\Delta}G(X_{T})=-k_{B}T\ln\frac{Z_{1}(X_{T})}{Z_{0}(X_{T})}={\Delta}G^{0}+W_{r}(\langle x_{r}\rangle_{1})-\langle f\rangle_{0}x_{m}+\frac{1}{2}k_{b}x_{m}^{2}+\Delta W_{h}~,
\label{s5}
\end{eqnarray}
where we used (\ref{r1b},\ref{r1c}) for the expressions of $Z_{0}$ and
$Z_{1}$ and we used the parameter $x_{m}$  defined as
the distance between the two states, $x_{m}=\langle x\rangle_{1}-\langle
x\rangle_{0}$. The functions $W_{r}$ and $\Delta W_{h}$ are given by
(\ref{rr1}) and (\ref{r5}). \\ The height of the barrier is given by the
difference of free energy between the F state an the transition state
that we will denote as $\sigma=t$ (averages taken when the molecule is in its transition state will be denoted by $\langle ...\rangle_{t}$). The transition state is located at
the point where the free energy landscape of the system depicted in
Fig.~\ref{f1} is maximum (Fig.~\ref{f3} (a)), and we define it as the
RNA state where the first $n^{*}$ bases are opened and the latter $N-n^{*}$ are
closed, $N$ being the total number of bases that form the RNA
molecule. Therefore the function $B(X_{T})$ is computed as the
free-energy difference between the folded state F and
the transition state, which are separated by a distance $x_{1}=\langle
x\rangle_{t}-\langle x\rangle_{0}$. This gives
\begin{eqnarray}
B(X_{T})=B^{0}+W_{r}(\langle x_{r}\rangle_{t})-\langle
f\rangle_{0}x_{1}+1/2k_{b}x_{1}^{2}+\Delta W_h^{t}~.
%=B^{1}-\langle f\rangle_{0}X_{1}+1/2k_{b}X_{1}^{2}.
\label{s6}
\end{eqnarray}
 The function $W_{r}$ is given by (\ref{rr1}) and $\Delta
W_h^{t}$ is the change in free energy of the handles when the RNA
molecule jumps from the F state to the transition state computed as:
\begin{eqnarray}
{\Delta}W_{h}^{t}=W_{h_{1}}(\langle x_{h_{1}}\rangle_{t})+W_{h_{2}}(\langle x_{h_{2}}\rangle_{t})-W_{h_{1}}(\langle x_{h_{1}}\rangle_{0})-W_{h_{2}}(\langle x_{h_{2}}\rangle_{0})~.
\label{rt5}
\end{eqnarray}
Then the
rates $k_{\rightarrow}$ and $k_{\leftarrow}$ associated to the
activated process can be written as:
\be
k_{\rightarrow}(X_{T})=k_{0}\exp[\beta(-B^{1} +\langle f\rangle_{0}x_{1}-1/2k_{b}x_{1}^{2})]~,
\label{rates1a}
\ee
\be
k_{\leftarrow}(X_{T})=k_{0}\exp[\beta(-B^{1}+\Delta G^{1}-\langle f\rangle_{1}x_{2}-1/2k_{b}x_{2}^{2})]~,
\label{rates1b}
\ee
\be
{\rm with} \;\;B^{1}=B^{0}+W_{r}(\langle x_{r}\rangle_{t})+\Delta W_h^{t}\;\;,\;\;\Delta G^{1}=\Delta G^{0}+W_{r}(\langle x_{r}\rangle_{1})+\Delta W_h\;,
\label{rates1c}
\ee
where we used (\ref{ssb1}), (\ref{s5}) and (\ref{s6}). The expression
for the rates (\ref{rates1a},\ref{rates1b}) are equivalent to the ones
obtained by Bell \cite{b26} but in the mixed ensemble. Note that the two
rates $k_{\rightarrow}(X_{T}),~k_{\leftarrow}(X_{T})$ satisfy the
detailed balance condition \eq{ss1b}.
 
\renewcommand{\theequation}{C-\arabic{equation}}
\setcounter{equation}{0}
\section{Demonstration of the equivalence between the  $N_{F}$ and  $N_{B}$ \label{NFR}}
Taking the expressions for the fractions $N_{F}$ and $N_{R}$ given by (\ref{s10a},\ref{s10b}) and integrating the left integral we get:
\begin{eqnarray}
N_{F}=1-\rho_{0}^{F}(y_{i},y_{f})+\int_{y_{i}}^{y_{f}}\frac{d\rho_{0}^{F}(y_{i},y)}{dy}\rho_{1}^{F}(y,y_{f})dy~,\nonumber\\
N_{R}=1-\rho_{1}^{R}(y_{f},y_{i})+\int_{y_{f}}^{y_{i}}\frac{d\rho_{1}^{R}(y_{f},y)}{dy}\rho_{0}^{R}(y,y_{i})dy~,
\label{CN1}
\end{eqnarray}
where $y$ denotes a generic control parameter.
Then using the equation for the evolutions of the probabilities $\rho_{\sigma}$ given by  (\ref{m5},\ref{m5b}) and for a symmetric perturbation protocol, $v_{F}(y)=\frac{dy}{dt}\Big{|}_{F}=-v_{R}(y)=\frac{dy}{dt}\Big{|}_{R}$, we obtain the following relation:
\begin{eqnarray}
\rho_{\sigma}^{F}(y',y)=\exp\Big{[}-\int_{y'}^{y}\frac{k_{\sigma \rightarrow \sigma '}(y'')}{v_{F}(y'')}dy''\Big{]}=\exp\Big{[}-\int_{y}^{y'}\frac{k_{\sigma \rightarrow \sigma '}(y'')}{v_{R}(y'')}dy''\Big{]}=\rho_{\sigma}^{R}(y,y').
\label{CN2}
\end{eqnarray}
We consider the expression for $N_{R}$ given by (\ref{CN1}) and we integrate by parts,
\begin{eqnarray}
N_{R}=1-\rho_{1}^{R}(y_{f},y_{i})-\rho_{0}^{R}(y_{f},y_{i})+\rho_{1}^{R}(y_{f},y_{i})-\int_{y_{f}}^{y_{i}}\frac{d\rho_{0}^{R}(y,y_{i})}{dy}\rho_{1}^{R}(y_{f},y)dy~.
\label{CN3}
\end{eqnarray}
Using the relation between the probabilities $\rho_{\sigma}$ for the forward and reverse process (\ref{CN2}), we obtain
\begin{eqnarray}
N_{R}=1-\rho_{0}^{F}(y_{i},y_{f})+\int_{y_{i}}^{y_{f}}\frac{d\rho_{0}^{F}(y_{i},y)}{dy}\rho_{1}^{F}(y,y_{f})dy=N_{F}~.
\label{CN4}
\end{eqnarray}

\renewcommand{\theequation}{D-\arabic{equation}}
\setcounter{equation}{0}
\section{Single domain RNA as a stick-slip process \label{app_singledomain}}
To address the case of a multidomain  molecule it is useful to focus first on the simpler case of a single domain molecule.  We consider an unfolding process without refolding ($k_{\leftarrow}=0$) characterized by an effective unfolding force-dependent rate $k_{\rightarrow}(f)$ (\ref{s111}). The distribution of breakage forces is given by \cite{b12} 
\begin{eqnarray}
P(f^{*})=\frac{k_{\rightarrow}(f^{*})e^{-\frac{k_{B}T}{r\tilde{x}_{1}}[k_{\rightarrow}(f^{*})-k_{\rightarrow}(0)]}}{r}~.
\label{m7}
\end{eqnarray}
Note that in order to the no refolding condition to be realistic there must be a limit force $f_{m}$, below which the distribution of breakage forces goes to zero. This lower limit $f_{m}$ arises because we are considering that there
 is a vanishing probability of jumping if the UF state is not thermodynamically
 stable, ${\Delta}G(f_{m})=0$. From this distribution one can
compute the mean value and the variance of the breakage force
 \cite{bb1,HumSza03}
\be
 \langle f^{*}\rangle=\int_{0}^{\infty}dfP(f)f=\frac{k_{B}T}{\tilde{x}_{1}}[-e^{a}Ei(-a)]~,
\label{m8}
\ee
where $a=\frac{k_{B}T}{r\tilde{x}_{1}}k_{\rightarrow}(0)$ and the function $Ei$ is the special elliptic function. By doing an expansion in the parameter $a$, that is much smaller than one (otherwise there is a finite probability of refolding), we obtain: 
\be
\langle f^{*}\rangle=\frac{k_{B}T}{\tilde{x}_{1}}\Big{[}\ln\Big{(}\frac{r\tilde{x}_{1}}{k_{B}Tk_{o}e^{-(\tilde{B}-1/2k_{b}\tilde{x}_{1}^{2})}}\Big{)}-\gamma\Big{]}+O(a)~.
\label{m10}
\ee 
where $\gamma$ is the Euler's constant, and
\be
\sigma_{f^*}^2=\langle f^{*2}\rangle-\langle f^{*}\rangle^{2}=(\frac{k_{B}T}{\tilde{x}_{1}})^{2}[{\pi}^{2}/6]+O(a)~.
\label{m12}
\ee 
 
Therefore by studying either the distribution of $f^*$ at fixed $r$ or the mean value or the variance of such distribution as a function of $r$ one can obtain information about the kinetic parameters doing a fit to (\ref{m7}), (\ref{m10}) or (\ref{m12}) respectively~\footnote{\label{foot19} The system under consideration verifies that the transition occurs close to the situation where the $k_{b}$ is much smaller than the other stiffness values, $k_{h_{1}},k_{h_{2}}$ and $k_{r}$. Beyond this regime, one should take into account the variability of $r$ with the force. And to get a more accurate result one should fit the data to the distribution of the breakage forces instead of the mean or variance of the breakage force as discussed in \cite{b31}.}.

 In Fig.~\ref{f12} we plot $\langle f^{*}\rangle$ and $\sigma_{f^*}^2$
as a function of $r$ obtained by pulling the system described in
Sec.~\ref{setup} with parameters given by Table~\ref{table1}. The kinetic parameters that characterize the RNA molecule that we consider here are given in Table~\ref{table3}.
\begin{table}[H] 
\begin{center}
\begin{tabular}{|c|c|c|c|c|}
\hline
$k_{0}\exp(-\beta B^{0})$ &$n^{*}$\\
\hline
$e^{-9}\approx 10^{-4}$& 2 \\ 
\hline  
\end{tabular}
\caption{Parameters that characterize the unfolding kinetics of the
RNA hairpin.\label{table3}}
\end{center}
\end{table}
We perform two kinds of simulation both using the condition of no refolding (\ref{m3}), but with the dynamics generated by different unfolding rates, the non-effective rates \eq{rates1a} and the effective rates (\ref{s111}) respectively:

{
\begin{itemize}
\item Non-effective rates: We consider the explicit dependence on
$X_{T}$ of the barrier $B(X_{T})$ that governs the unfolding kinetics, see \eq{s6}.

\item
Effective rates: We use the unfolding effective rate (\ref{s111}) in
order to generate the unfolding dynamics where we neglect the dependence on $X_{T}$ of $x_{1}$ and $B^{1}$ as given by \eq{rates1c}. For kinetics processes with
barriers quite insensitive 
to the force (or $X_T$) this seems to be a
reasonable approximation. The effective model is also the one we use to do the
analytical computations.

\end{itemize}

The comparison between both simulations allows us to see how big are the
difference between both models, and how far the analytical results
are from the non-effective model. In Fig.~\ref{f12} (b)
we show $\sigma_{f^*}^2$ for both kinds of simulations for a broad range
of values of $r$. The non-effective simulation gives fluctuations $\sigma_{f^*}^2$
that decrease when $r$ increases, instead of being constant as the
effective model predicts. This effect comes from the dependence of
$\tilde{x}_{1}$ on $f$ (or $X_{T}$). On the other hand we can see that
when $r$ approaches zero, fluctuations disappear, because the domain
is always opened at zero force~\footnote{In the simulation there is no 
restriction for the breakage force, hence $f^*$ can be smaller than $f_m$.
When $r$ goes to zero the breakage force too, because 
the molecule always opens if we wait long enough (and does not close anymore as $k_{\leftarrow}=0$).}. 
 However, we should note that in the
latter regime ($r$ going to zero) the no-refolding approximation becomes
invalid, because the distribution $P(f^{*})$ do not vanish for $f^{*}<f_{m}$. Nevertheless we see that for the range of interest $r\approx
1-50\rm{pN/s}$ both simulations agree pretty well either for the
$\sigma_{f^*}^2$ as for the $\langle f^{*}\rangle$. We can conclude that
the effective dynamics reproduces well the non-effective one.
\begin{figure}[H]
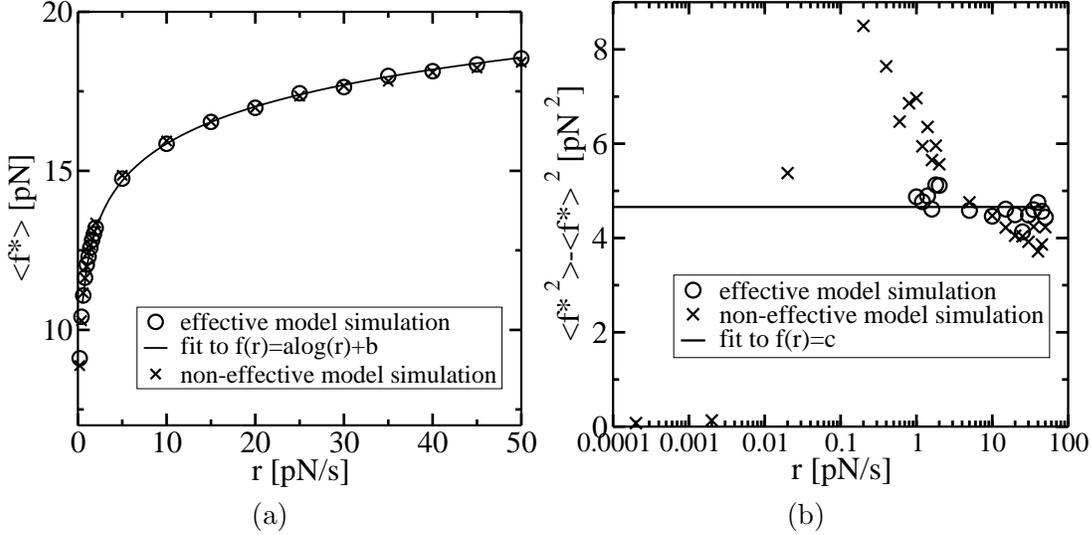

\begin{minipage}{7cm}
\begin{center}
\includegraphics[height=6.5cm]{ff.eps}\\
(a)
\end{center}
\end{minipage}
\begin{minipage}{7cm}
\begin{center}
\includegraphics[height=6.5cm]{fff.eps}\\
(b)
\end{center}
\end{minipage}
\caption{\small{Comparison between the effective and non-effective rates. 
Results for 3000 pulling trajectories for a domain
characterized by the kinetics parameters given in
Table~\ref{table3}. For the effective kinetics parameters we use
$\tilde{x}_{1}=2.5\rm nm$ and $\tilde{B}\ln(k_{o})==8.5k_{B}T$. In (a) it is shown the mean breakage force as a
function of the
pulling rate. The straight line is the best fit to a function $y=a\ln (x)+b$,
obtaining $\tilde{x}_{1}=2.47\pm 0.03\rm nm$, $\tilde{B}\ln(k_{o})=8.3\pm
0.2k_{B}T$. In (b) we represent the variance in breakage force as a
function of the pulling rate. The straight line is the best fit to a
constant $y=C$ for the data with $r>1\rm{pN/s}$, obtaining
$\tilde{x}_{1}=2.45\pm 0.06\rm nm$. }}
\label{f12}
\end{figure}

Fitting the data obtained from the simulation for $\langle f^{*}(r)\rangle$ to (\ref{m10}), and for $\sigma_{f^*}^2$ to (\ref{m12}), we get accurate results for both parameters, $\tilde{x}_{1}$ and $\tilde{B}\ln(k_{o})$.

\renewcommand{\theequation}{E-\arabic{equation}}
\setcounter{equation}{0}
\section{Computation of the distribution of probability of breakage force in regime 3\label{PF3}}
According to (\ref{m7}) the distribution of  breakage forces for the $i$th domain conditioned to a given value of $f_{i}^{s}$ (with $f_{i}^{s}=f_{i-1}^{*}-k_{b}\tilde{x}_{m}^{(i-1)}$), $\rho(f_{i}^{*}|f_{i}^{s})$, is:

\be
\rho(f_{i}^{*}|f_{i}^{s})=\frac{k_{\rightarrow}^{(i)}(f_{i}^{*})e^{\frac{k_{B}T}{r\tilde{x}_{1}^{(i)}}[k_{\rightarrow}^{(i)}(f_{i}^{*})-k_{\rightarrow}^{(i)}(f_{i}^{s})]}}{r}\theta_{H}(f_{i}^{*}-f_{i}^{s})~,
\label{m14}
\ee
where the parameters and functions with index $i$ refer to the domain $i$. The $\theta_{H}$ is the Heaviside function, $\theta_{H}(x)=1$ only if $x>0$ otherwise $\theta_{H}(x)=0$. Assuming $\tilde{x}_{m}^{(i-1)}$ as a constant parameter we derive the breakage force distribution $P(f_{i}^{*})$ averaging (\ref{m14}) over the distribution of the  breakage forces of the previous domain $i$-1, $P(f_{i-1}^{*})$. To get $P(f_{i-1}^{*})$ one has to average over the distribution of the  breakage forces of the domain $i$-2, and so on. This leads to the following recurrence formula:
\be
P(f_{i}^{*})=C\int_{-\infty}^{\infty}\rho(f_{i}^{*}|f_{i}^{s})\prod_{k=2}^{i-1}df_{k}^{*}\rho(f_{k}^{*}|f^{s}_{k})df_{1}^{*}P(f_{1}^{*})~,
\label{m16}
\ee
where $C$ is a normalization factor. We will consider the case where the distribution of breakage force for the domain $i$-1 is not modified by the previous one, either because it is the first domain, or because the typical value of $f_{i-1}^{*}$ is higher than  all the  previous ones $f_{k}^{*}$ with $k<i-1$. Then the distribution of $f_{i-1}^{*}$ is computed as in the case of a single barrier (\ref{m7}) and (\ref{m16}) reduces to,
\be
P(f_{i}^{*})=C\int_{-\infty}^{\infty}\rho(f_{i}^{*}|f_{i}^{s})P(f_{i-1}^{*})df_{i-1}^{*}~.
\label{m16b}
\ee
The integral in \eq{m16b} is not analytically solvable. 
Then we expand in power series the exponential $\exp\Big{[}{-\frac{k_{B}T}{r\tilde{x}_{1}^{(i)}}[k_{\rightarrow}^{(i)}(f_{i}^{s})]}\Big{]}$ in \eq{m14} and substituting in \eq{m16b} we obtain:

\begin{eqnarray}
P(f_{i}^{*})=C\frac{k_{\rightarrow}^{(i)}(f_{i}^{*})e^{\frac{k_{B}T}{r\tilde{x}_{1}^{(i)}}k_{\rightarrow}^{(i)}(f_{i}^{*})}}{r}[\sum_{j=0}^{\infty}\frac{A_{i}^{j}}{j!}]~,
\label{m17}
\end{eqnarray}
where 
\begin{eqnarray}
A_{i}^{j}&=&\int_{-\infty}^{f_{i}^{*}+k_{b}\tilde{x}_{m}^{(i-1)}}C_{i}P(f_{i-1}^{*})df_{i-1}^{*}=\frac{[(k_{B}T/r\tilde{x}_{1}^{(i)})k_{\rightarrow}^{(i)}(0)\exp(-\frac{k_{b}\tilde{x}_{m}^{(i-1)}\tilde{x}_{1}^{(i)}}{k_{B}T})]^{j}}{[(k_{B}T/r\tilde{x}_{1}^{(i-1)})k_{\rightarrow}^{(i-1)}(0)]^{\tilde{x}_{1}^{(i)}j/\tilde{x}_{1}^{(i-1)}}}\nonumber\\
& &\times \gamma(j\tilde{x_1}^{(i)}/\tilde{x_1}^{(i-1)}+1,\frac{k_{B}T}{r\tilde{x_1}^{(i-1)}}k_{\rightarrow}^{(i)}(f_{i}^{*}+k_{b}\tilde{x}_{m}^{(i-1)}))~.
\label{m18}
\end{eqnarray}
The $\gamma(x,y)$ is the incomplete gamma function. For the regime 3
the series can be truncated, because the moments of $C_{i}$,
$A_{i}^{j}$, are not big and the series fastly converges.


\begin{thebibliography}{00}


\bibitem{b1} J.A. Doudna and T.R. Cech, (2002) Nature {\bf 418}, 222-228
\bibitem{b2} P.B. Moore and T.A. Steitz, (2002) Nature {\bf 418}, 229-235
\bibitem{b3} C. Bustamante, J. Macosko and G. Wuite, (2000) Nature Reviews, Molecullar Cell Biology {\bf 1}, 130-136 
\bibitem{b4} S. Smith, L. Finzi and C. Bustamante, (1992) Science {\bf 258}, 1122-1126.
\bibitem{b5} S. Smith, Y. Cui and C. Bustamante, (1996) Science {\bf 271}, 795-799
\bibitem{b6} P. Cluzel, A. Lebrun, C. Heller, R. Lavery, J.L. Viovy, D. Chatenay, D. and F. Caron, (1996) Science {\bf 271}, 792-794.
\bibitem{b7} B. Essevaz-Roulet, U. Bockelmann and F. Heslot, (1997)  Proc. Natl. Acad. Sci. USA {\bf 94}, 11935-11940.
\bibitem{b8}  R. Russell, X.W. Zhuang, H.P. Babcock, I.S. Millet, S. Doniach, S. Chu and D. Herschlag, (2002) Proc. Natl. Acad. Sci. USA {\bf 99}, 155-160
\bibitem{b9} X.W. Zhuang, H. Kim, M.J.B. Pereira, H.P. Babcock, N.G. Walter and S. Chu, (2002) Science {\bf 296}, 1473-1476 
\bibitem{b10} D. Liphardt, B. Onoa, S. Smith, I. Tinoco and C. Bustamante, (2001) Science {\bf 292}, 733-737.
\bibitem{b11} B. Onoa, D. Dumont, J. Liphardt, S. Smith, I. Tinoco and C. Bustamante, (2003) Science {\bf 299}, 1892-1895 
\bibitem{b12} E. Evans and K. Richie, (1997) Biophys. J {\bf 72}, 1541-1555
\bibitem{b13} E. Evans and K. Richie,  (1999) Biophys. J {\bf 76}, 2439-2447

\bibitem{b13bi} S. Smith, Y. Cui and C. Bustamante, (2003), Methods in Enzymology {\bf 361}, 134-162. 
\bibitem{b13bibi} D. Keller, D. Swigon and C. Bustamante, (2003)  Biophys. J. {\bf 84}, 733-738 

\bibitem{b14} U. Gerland, R. Bundschuh and T. Hwa, (2003) Biophys. J. {\bf 84}, 2831-2840.  
\bibitem{b15} U. Gerland, R. Bundschuh and T. Hwa, (2001) Biophys. J. {\bf 81}, 1324-1332.  

\bibitem{b16} F. Ritort, C. Bustamante and I. Tinoco, (2002)  Proc. Natl. Acad. Sci. USA {\bf 99}, 13544-13548.

\bibitem{b17} J.M. Fernandez, S. Chu and A.F. Oberhauser, (2001) Science {\bf 292}, 653-654.

\bibitem{b18} V. Mu\~{n}oz, P.A. Thompson, J. Hofrichter and W.A. Eaton, (1997) Nature {\bf 390}, 196. 
\bibitem{b19} G. Bokinsky, D. Rueda, V.K. Misra, A. Gordus, M.M. Rhodes, H.P. Babcock, N.G. Walter and  X. Zhuang, (2003)  Proc. Natl. Acad. Sci. USA {\bf 100}, 9302-9307.
\bibitem{b20} X.W. Zhuang,, T. Ha, H.D. Kim, T. Centner, S. Labeit and S. Chu, (2000) Proc. Natl. Acad. Sci. USA {\bf 97}, 14241-14244
\bibitem{HumSza03} G. Hummer and A. Szabo, (2003)  Biophys. J. {\bf 85}, 5-15 
\bibitem{CocMonMar} S. Cocco, R. Monasson and J. Marko, (2003) Eur. Phys. J. E {\bf 10}, 153
\bibitem{Marinarietal} E. Marinari, A. Pagnani and F. Ricci-Tersenghi, (2002) Phys. Rev. E {\bf 65}, 041919 
\bibitem{b21} I. Tinoco and C. Bustamante, (2002) Biophysical Chemistry {\bf 102}, 513-533. 
\bibitem{b22} P.J. Flory, (1969) Statistical mechanics of chain molecules, appendix G, Oxford University Press (NY).  
\bibitem{b23} J.F. Marko and E.D Siggia, (1995) Macromolecules {\bf 28}, 8759-8770.
\bibitem{b24} C. Bustamante, J.F. Marko, E.D. Siggia, and S. Smith, (1994) Science {\bf 265}, 1599-1600.
\bibitem{b25} C. Bouchiat, M. Wang, J. Allemand, T. Strick, S. Block and V. Croquette, (1999) Biophys. J. {\bf 76}, 409-413.
\bibitem{b26} I.G. Bell,  (1978) Science {\bf 200}, 618-627
\bibitem{bb26} A. Imparato and L. Peliti, (2004) EPJ, PREPRINT
\bibitem{b27} P. Zarrinkar and J. Williamson, (1994) Science {\bf 265}, 918-923.

\bibitem{b28} X.W. Fang, P. Thiyagarajan, T.R. Sosnick and T. Pan, (2002) Proc. Natl. Acad. Sci. USA {\bf 99}, 8518-8523.

\bibitem{b29} R. Russell, I. Millet, M. Tate, L. Kwok, B. Nakatani , S. Gruner, S. Mochrie, V. Pande, S. Doniach, D. Herschlag and L. Pollak, (2002) Proc. Natl. Acad. Sci. USA {\bf 99}, 4266-4271.
\bibitem{b30} X.W. Zuang, L.E. Bartley, H.P. Babcock, R. Russell, T.J. Ha, D. Herschlag and S. Chu, (2000) Science {\bf 288}, 2048-2051  
\bibitem{b31} C. Friedsam, A.K. Wehle, F. k\"{u}hner, and H.E. Gaub, (2003) J. Phys: Condens. Matter {\bf 15}, S1709-S1723
\bibitem{bb1} C. Gerley, J.C. Voegel, P. Schaaf, B. Senger, M. Maaloum, J.K.H. Horber and J. Hemmerle, (2000) Proc. Natl. Acad. Sci. USA {\bf 97}, 10802-10807.
\bibitem{b32} S. Harlepp, T. Marchal, J. Robert, J-F. L\'eger, A. Xayaphoummine, H. Isambert and D. Chatenay, (2003) arXiv:physics/0309063 v1 
	
\bibitem{Jarzynski}  J. Liphardt, S. Dumont, S.B. Smith, I. Tinoco
Jr. and C. Bustamante, (2002) Science {\bf 296}, 1832-1835.

\bibitem{Ritort} F. Ritort, (2003) Poincar\'e Seminar {\bf 2},
195-229. Preprint arXiv:cond-mat/0401311

\bibitem{Chandler} P. G. Bolhuis, D. Chandler, C. Dellago and
P. L. Geissler, (2002) Ann. Rev. Phys. Chem. {\bf 53}, 291-318.









\end{thebibliography}
\end{document}